\documentclass[11pt,preprint]{aastex} 
\newcommand{\gsim}{\mbox{$\stackrel {>}{_{\sim}}$}} 
\newcommand{\lsim}{\mbox{$\stackrel {<}{_{\sim}}$}} 
 
\slugcomment{Accepted to ApJ}
 
\shorttitle{Chemistry in IC 342} 
\shortauthors{Meier \& Turner} 
 
\received{2004 July 29}

\begin{document} 
 
\title{Spatially Resolved Chemistry in Nearby Galaxies I. The Center
of IC 342}

\author{David S. Meier}
\affil{Department of Astronomy, University of Illinois, Urbana-Champaign, 
1002 W. Green St., Urbana, IL 61801}
\email{meierd@astro.uiuc.edu} 
\and
\author{Jean L. Turner}  
\affil{Department of Physics and Astronomy, UCLA, Los Angeles, CA 90095--1547}
\email{turner@astro.ucla.edu} 

\begin{abstract} 
We have imaged emission from the millimeter lines of eight
molecules---C$_2$H, C$^{34}$S, N$_2$H$^+$, CH$_3$OH, HNCO, HNC,
HC$_3$N, and SO---in the central half kiloparsec of the nearby spiral
galaxy IC 342. The 5\arcsec\ ($\sim 50$ pc) resolution images were
made with the Owens Valley Millimeter Array (OVRO).  Using these and
previously published CO and HCN images we obtain a picture of the
chemistry within the nuclear region on the sizescales of individual
giant molecular clouds.  Bright emission is detected from all but SO.
There are marked differences in morphology for the different
molecules.  A principal component analysis is performed to quantify
similarities and differences among the images.  This analysis reveals
that while all molecules are to zeroth order correlated, that is, they
are all found in dense molecular clouds, there are three distinct
groups of molecules distinguished by the location of their emission
within the nuclear region. N$_2$H$^+$, C$^{18}$O, HNC and HCN are
widespread and bright, good overall tracers of dense molecular gas.
C$_2$H and C$^{34}$S, tracers of Photo-Dissociation Region chemistry,
originate exclusively from the central 50-100 pc region, where
radiation fields are high. The third group of molecules, CH$_3$OH and
HNCO, correlates well with the expected locations of bar-induced
orbital shocks.  The good correlation of HNCO with the established
shock tracer molecule CH$_3$OH is evidence that this molecule, whose
chemistry has been uncertain, is indeed produced by processing of
grains.  HC$_{3}$N is observed to correlate tightly with 3mm continuum
emission, demonstrating that the young starbursts are the sites of the
warmest and densest molecular gas.  We compare our HNC images with the
HCN images of Downes et al. (1992) to produce the first high
resolution, extragalactic HCN/HNC map: the HNC/HCN ratio is near unity
across the nucleus and the correlation of both of these gas tracers
with the star formation is excellent.  The ratio exhibits no obvious
correlation with gas temperature or star formation strength.

\end{abstract} 
\keywords{galaxies: individual(IC 342)  ---  galaxies: starburst
--- galaxies: ISM --- radio lines: galaxies --- astrochemistry}
 
\section{Introduction \label{intro}} 

The bright and abundant molecule CO has dominated the study of
molecular clouds in external galaxies.  The millimeter-wave
transitions of CO and its isotopomers are powerful probes of diffuse
molecular gas \citep[eg.][]{YS91}.  However, emission from the
optically thick and easily excited CO tends to favor the warmer,
radiatively lit and diffuse surface layers of clouds \citetext{eg.,
\citealp{THH93}; Meier, Turner \& Hurt 2000, hereafter
\citeauthor{MTH00}}.  Since dense gas is the component most closely
connected to star formation \citep*[eg.,][]{GS04}, to understand the
links between molecular clouds and star formation in different
galactic environments, we need to study tracer molecules appropriate
to a cooler, dense component that may not be well traced by CO.

Each molecular line traces a distinct regime of density and
temperature within a molecular cloud, while different molecules can
trace different gas chemistries \citep*[see][for a review of Galactic
astrochemistry]{VB98}.  Emission from high density tracers such as
HCN, HCO$^{+}$ and CS show that large amounts of dense ($\gsim ~
10^{4}$ cm$^{-3}$) molecular gas are present in the centers of
galaxies \citep*[eg.][]{MH89,MHWH89,NNJ89,NJHTM92,SDR92,HB93}, and
that the dense molecular interstellar medium (ISM) can vary
significantly on scales of a few tens of parsecs
\citep*[eg.][]{Dow92,BS93,HB97,SFB97,SNKN02}.

Numerous molecules have been detected in nearby starbursts
\citep*[eg.,][]{HJMMS87,HSM88,NHJM91,MHWSW91,PB92,SZ95,
MHC95,MMMGH03,UGFMR04,WHCWHMM04}.  These studies find that the
chemical differentiation seen within Galactic molecular clouds
survives to scales of hundreds of parsecs, and even to galaxy-wide
differences.  However, most of these observations have been done with
single dish telescopes, at resolutions of $\sim$30\arcsec, or a few
hundred pc on the galaxy, and thus average many giant molecular clouds
(GMCs) together into one beam.

With millimeter interferometers it is now possible to resolve
individual GMCs in the nearest galaxies with sufficient sensitivity to
allow the study of selected chemical species
\citep*[eg.,][]{GMFN00,GMFN01,GMFUN02}.  Because of the low ($\sim
50-100$ pc) physical resolution, extragalactic observations are
insensitive to small-scale ($\sim$tenths of pc) chemistry typically
studied in the Galaxy, but such maps can resolve the chemical
properties operating over the bulk of a GMC and between individual
GMCs.  These observations can provide insights regarding the extent to
which distinct large-scale properties, such as starbursts, shocks,
bars, spiral arms, tidal forces and AGN influence the chemistry of
molecular clouds and how these influences are transmitted from the
GMC-scale to the galaxy as a whole.

We have surveyed the nearby Scd galaxy, IC 342, in transitions from
eight astrochemically important molecules. C$_{2}$H, HNCO, HNC,
HC$_{3}$N, N$_{2}$H$^{+}$, C$^{34}$S, CH$_{3}$OH, and SO were imaged
at $\sim$50 pc resolution with the Owens Valley Millimeter Array
(OVRO).  These are the first published interferometer maps of these
lines in an external galaxy.  IC 342 is the nearest (D $\sim$ 2
Mpc)\footnote{Due to its location behind the Galactic plane, IC 342's
distance is still a matter of debate. Recent studies have argued for a
distance of $\sim$3 Mpc \citep*[eg.,][]{SCH02,KSDG03}, but we adopt
the shorter distance to keep consistent with our previous work.},
gas-rich spiral with active star formation in its nucleus
\citep*[][]{Bec80,TH83}, and the first galaxy to be mapped with
millimeter interferometers \citep{L84}.  There is widespread
resolvable molecular gas distributed in both dense clouds, with masses
similar to SgrB2, and a diffuse medium about the size of the central
molecular zone \citep{MS96} in the Milky Way.  Detailed knowledge of
the H$_{2}$ column densities, excitation temperatures and densities in
IC 342 exists from CO and its isotopomers as a basis for comparison
with these new lines \citetext{\citealp{I90,Dow92,TH92,WITHL93,THH93};
\citeauthor{MTH00}; Meier \& Turner 2001, hereafter
\citeauthor{MT01}}.

\section{Observations \label{obs}} 
 
We observed eight lines at 3mm and one line at 1mm with OVRO between
1997 October 22 and 2001 April 07 (Table \ref{ObsT}).  The
interferometer consisted of six 10.4 meter antennas with SIS receivers
\citep{OVRO91, OVRO94}.  All transitions were observed in the C and L
array configurations except for the SO, which was observed in L and H.

Table \ref{ObsT} lists the observed lines along with the observational
parameters.  The transitions were selected based on the criteria that
(1) they are bright in SgrB2 (T$_{a}\gsim$1 K in the \citet[][]{T89}
spectral line survey), (2) they sample a selection of different types
of chemistry and (3) they maximize the number of species OVRO can
observe simultaneously.  Table \ref{MolP} lists molecular parameters
for the transitions.  The nine spectral lines were observed in three
sets of spectrometer configurations.  C$_{2}$H, HNCO, HNC and
HC$_{3}$N were observed as a group, as were CH$_{3}$OH, C$^{34}$S and
N$_{2}$H$^{+}$, and the two SO transitions were observed with
C$^{18}$O (\citeauthor[][]{MT01}).  Each group has the same
instrumental configuration, phase center, and weather.  Data were
calibrated using the MMA package.  Phase calibration was done by
observing the point source 0224+671 every 20 minutes.  Absolute flux
calibration is based on observations of Neptune or Uranus and with
3C273, 3C84, 3C454.3 and 3C345 as supplementary flux calibrators.
Based on the derived fluxes and flux histories of these secondary flux
calibrators we estimate that the absolute fluxes are good to 10 - 15\%
for the 3 mm data and 20 - 25\% for the 1 mm data (SO).  The lines of
SO($2_{3}-1_{2}$) and SO($6_{5}-5_{4}$) were not detected.

Both robustly weighted maps with resolutions of $\sim5-6{''}$ and
uniformly weighted maps with $\sim4^{''}$ resolution were produced.
The maps are not primary beam corrected.  Data reduction was done with
the NRAO AIPS.  In making the integrated intensity maps emission
greater than 1.2$\sigma$ was included.  Continuum emission has not
been subtracted from the maps since the 3 mm continuum peak is below
1$\sigma$.

The (u,v) coverages imply that emission on scales larger than $\sim
50^{''}$ is resolved out.  To estimate the amount of extended flux
missing from the images, each map was compared with its single-dish
spectrum.  Though somewhat uncertain due to the low S/N of some of the
single-dish detections
\citep*[][]{HSM88,MH91b,NHJM91,MHC95,HHMBWM95,HMH97}, all species are
consistent with no flux being resolved out, except possibly HC$_{3}$N.
This is consistent with what is found for $^{13}$CO and C$^{18}$O
(\citeauthor[][]{MTH00,MT01}) towards IC 342, and is expected for
these dense gas tracers.  In the case of HC$_{3}$N, the interferometer
HC$_{3}$N flux is $\sim$30 \% of the claimed tentative detection by
IRAM \citep*[][]{HSM88}, but given its very high dipole moment it is
not expected to be extended on $\sim$50 pc scales.  This implies that
the claimed single-dish brightness for HC$_{3}$N was overestimated,
and not that the interferometer resolves out significant flux.

\section{IC 342 and its Molecules}

\subsection{A Sketch of the Nucleus of IC 342 \label{morph_gen}} 

\begin{figure}
\epsscale{0.5}
\plotone{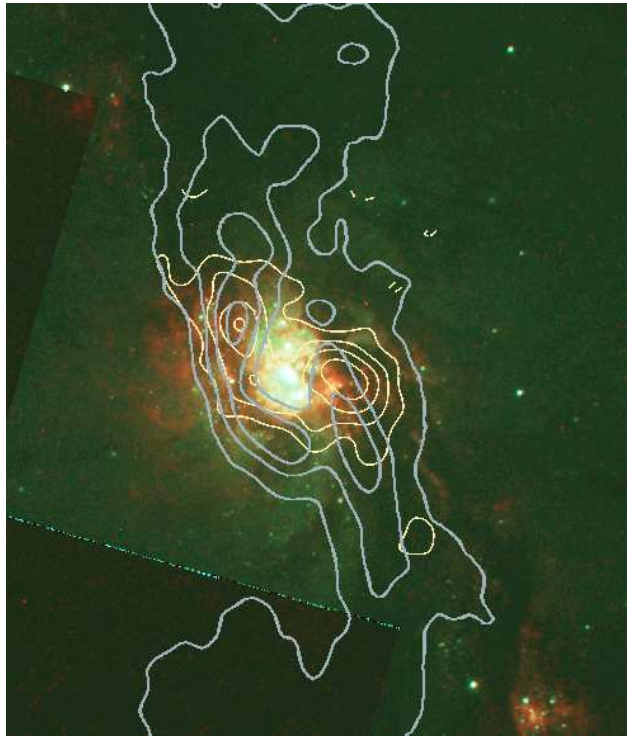}
\caption{A color HST composite of IC 342's nucleus.  Green/Blue is
F555 (V band) and red is H$\alpha$ + continuum.  The blue contours are
$^{12}$CO(1-0) and the yellow contours are 3 mm \citetext{see
\citeauthor{MT01}}. \label{hst} }
\end{figure}

We have a basic understanding of the small-scale molecular structure
of IC 342's nucleus from studies of CO, its isotopomers and HCN
\citetext{\citealp{I90,TH92}; \citeauthor{MTH00,MT01};
\citealp{SGKK01,MHWPH03}}.  A CO(1-0) map of IC 342 \citep[][]{LTH94}
is shown atop an HST image of the nucleus in Figure \ref{hst}.  Within
the central 300 pc (30\arcsec ) two molecular arms extend inward
\citep*[][]{I90}, terminating in a central ring of dense gas
\citep[][]{Dow92}. The total mass of molecular gas within the central
kpc is $\sim 4 \times 10^7~M_\odot$.  Orbital timescales here are a
few x $10^7$ yrs. The gas is on pronounced oval orbits, with an
estimated radial drift into the nuclear region due to tidal torquing
of $\rm \sim 0.1~ M_\odot~yr^{-1}$ \citep*[][]{TH92}, which is also
the approximate rate of current star formation in the radio/IR
source. The central molecular ring surrounds a nuclear star cluster
estimated to be 6-60 Myr in age \citep{BFG97}.  The star cluster
coincides with a central ``trough'' of molecular gas, the hole in the
molecular ring.

Five prominent GMCs with masses of $\rm \sim 10^{6}~M_{\odot}$ are
found within the molecular ring and arms.  These clouds have masses
slightly less than the Sgr B2 cloud in the Galactic Center. GMCs B and
C (adopting the nomenclature of \citealt{Dow92}) are located where the
incoming molecular arms meet the ring. These clouds coincide with two
young ($\sim$ few Myr old) large star-forming regions. GMC B is near
the dominant of the two IR/radio star-forming regions, which has a
luminosity of $\rm L_{OB}\sim 10^8~L_\odot$ \citep{Bec80,TH83}
corresponding to an estimated 500 O stars.  Cloud C appears somewhat
warmer than B in the highest transitions of CO \citep[][]{HSGRGH91},
but a careful analysis at high resolution in the lower transitions GMC
B is actually the warmest location in the nucleus (\citeauthor{MT01}).
GMC A is closest in projection to the nuclear star cluster and
dynamical center.  GMC A has similar CO(1-0) and HCN(1-0) properties
to B and C, but much weaker star formation.  GMC D, along the northern
arm, is also not a site of strong star formation. In this region large
gas streaming motions are observed \citep{TH92} along the arms. The
resulting shear could slow star formation, although there is some star
formation indicated by H$\alpha$ downstream from the molecular arms.

\subsection{Overview of the Molecules of IC 342\label{tempover}}

\begin{figure}
\epsscale{0.6}
\plotone{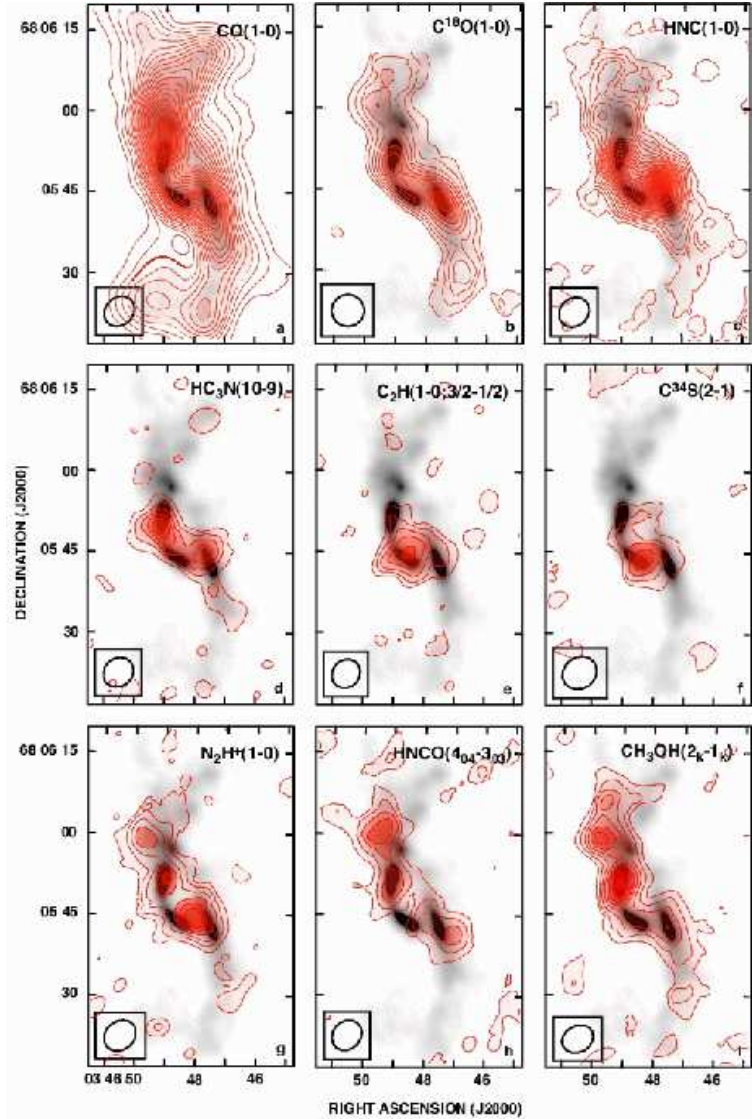}
\caption{The integrated intensities of nine tracer molecules observed
towards the nucleus of IC 342.  Each map is contoured in steps of the
2$\sigma$ times the RMS intensity in each map, except for
$^{12}$CO(1-0).  (a) The $^{12}$CO(1-0) transition smoothed to
$5''\times 4.''5$ resolution and contoured in steps of 7.7 K km
s$^{-1}$.  Locations of five major GMCs \citep*[][]{Dow92} are
displayed.  (b) The C$^{18}$O(1-0) transition contoured in steps of
1.4 K km s$^{-1}$ for $6.''0$ resolution, (c) the HNC(1-0) transition
in steps of 1.5 K km s$^{-1}$, (d) the HC$_{3}$N(10-9) transition in
steps of 1.5 K km s$^{-1}$, (e) the C$_{2}$H N(J,F)=
1($\frac{3}{2}$,2)-0($\frac{1}{2}$,1) transition in steps of 3.0 K km
s$^{-1}$, (f) the C$^{34}$S(2-1) transition in steps of 1.8 K km
s$^{-1}$, (g) the N$_{2}$H$^{+}$(1-0) transition in steps of 1.6 K km
s$^{-1}$, (h) the HNCO($4_{04}-3_{03}$) transition in steps of 2.1 K
km s$^{-1}$, and (i) the CH$_{3}$OH($2_{k}-1_{k}$) transition in steps
of 1.8 K km s$^{-1}$.  The greyscale seen in all frames is of
$^{12}$CO(1-0) at full resolution \citetext{$2.''7\times 2.''2$; see
\citeauthor{MTH00}} for comparison.\label{IntI} }
\end{figure}

\begin{figure}
\epsscale{0.9}
\plotone{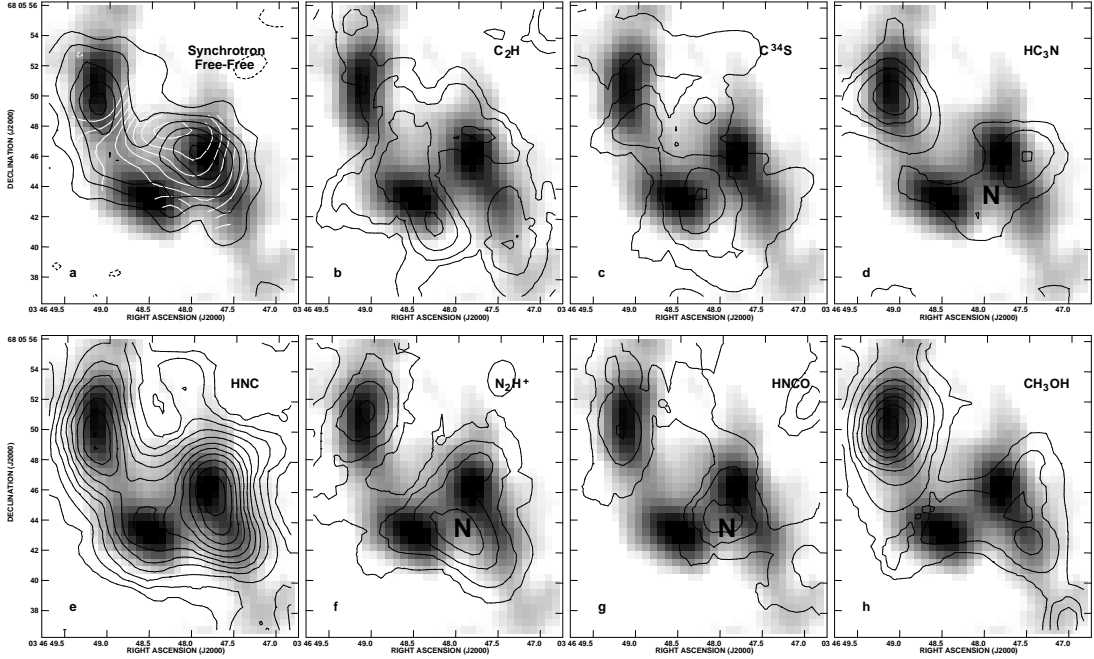}
\caption{Integrated intensities of the seven tracer molecules observed
in the central portion of IC 342's nucleus, weighted to optimize
spatial resolution at the expense of sensitivity.  Each map is
contoured at $\sim 2\sigma$ times the RMS noise level in each map
except for 6 cm synchrotron.  Each transition is overlaid the
greyscale image of HCN(1-0) kindly provided by D. Downes
\citep*[][]{Dow92}.  The HCN(1-0) greyscale ranges from 10 - 82 K km
s$^{-1}$ for a resolution of 2.$^{''}$8x2.$^{''}$7; pa = -180.  The
location of the potential nitrogen core is labelled by ``N'' in the
corresponding panels.  (a) The 3mm continuum emission \citetext{black
contours; \citeauthor{MT01}} in steps of 1 mJy beam$^{-1}$ for a
resolution of 3.$^{''}$3x3.$^{''}$0; -44.3$^{o}$, together with the 6
cm synchrotron flux (white contours) at the same resolution and
contour increment.  The 6 cm synchrotron emission is obtained by
assuming the 3 mm continuum emission (\citeauthor{MT01}) is all
thermal free-free and extrapolating this flux to 6 cm, and removing it
from the 6 cm continuum flux of \citet*[][]{TH83}.  (b) C$_{2}$H
N(J,F)= 1($\frac{3}{2}$,2)-0($\frac{1}{2}$,1) transition in steps of
3.6 K km s$^{-1}$ for 4.$^{''}$5x4.$^{''}$0; -21.7$^{o}$ resolution.
(c) C$^{34}$S(2-1) transition in steps of 3.1 K km s$^{-1}$ for
5.$^{''}$0x4.$^{''}$3; -51.7$^{o}$ resolution.  (d) HC$_{3}$N(10-9)
transition in steps of 2.8 K km s$^{-1}$ for 4.$^{''}$6x4.$^{''}$1;
-30.2$^{o}$ resolution. (e) HNC(1-0) transition in steps of 3.3 K km
s$^{-1}$ for 4.$^{''}$5x4.$^{''}$0; -24.6$^{o}$ resolution.  (f)
N$_{2}$H$^{+}$(1-0) transition in steps of 2.9 K km s$^{-1}$ for
4.$^{''}$7x4.$^{''}$1; -37.3$^{o}$ resolution.  (g)
HNCO($4_{04}-3_{03}$) transition in steps of 4.2 K km s$^{-1}$ for
4.$^{''}$7x4.$^{''}$0; -22.3$^{o}$ resolution.  (h)
CH$_{3}$OH($2_{k}-1_{k}$) transition in steps of 2.5 K km s$^{-1}$ for
4.$^{''}$6x4.$^{''}$0; -51.7$^{o}$ resolution.  \label{hr_mol}}
\end{figure}

\begin{figure}
\epsscale{0.75}
\plotone{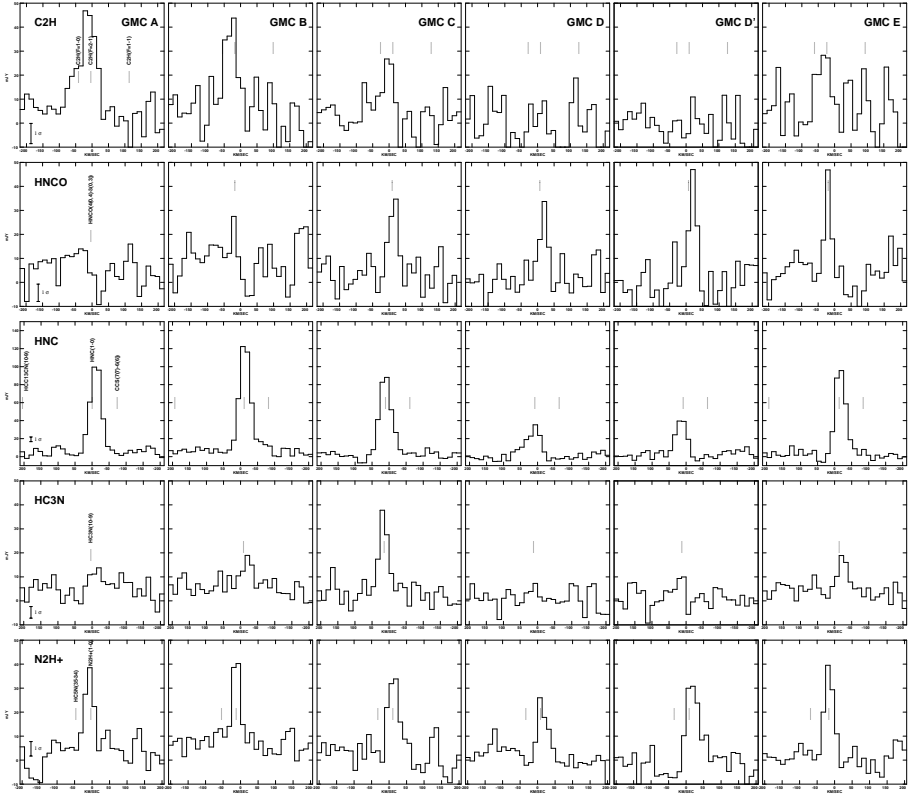}
\caption{Spectra from five of the observed transition sampled at the
locations of the five major GMCs \citep*[eg.][]{Dow92}.  Spectra are
summed of a 6$^{''}$ box centered on the fitted locations of each GMC.
Positions for each GMC are taken from high resolution C$^{18}$O
observations \citetext{Table 3 of \citeauthor{MT01}}.  GMC A is at the
far left of the each set of spectra, extending to GMC E is at the far
right.  Note that the sampled locations do not necessarily align
precisely with the peaks of the tracer molecules.  In the first plane
of the each set, any spectral line $\rm T_{mb}\gsim$ 0.1 K in the
\citet*[][]{T89} SgrB2 line survey that falls within the observed
bandwidth is labeled.  For the remaining four panels these lines are
marked by tick marks shifted in velocity to match IC 342's
$^{12}$CO(1-0) velocity at that location.  In all cases the zero
velocity corresponds to $v_{LSR}=35$ km s$^{-1}$ of the transition(the
brightest component if there is unresolved hyperfine structure).  Also
included in the first plane is the 1$\sigma$ errorbar, determined from
the RMS in an individual linefree channel. \label{spec1}}
\end{figure}

Figure \ref{IntI} displays the robustly-weighted integrated intensity
maps for the seven detected lines plus smoothed maps of previously
published $^{12}$CO(1-0) and C$^{18}$O(1-0) (\citeauthor{MTH00,MT01}).
Each map is overlaid on a greyscale image of $^{12}$CO(1-0).  Figure
\ref{hr_mol} displays the higher resolution uniformly weighted maps
for the central ring region overlaid on the greyscale HCN(1-0) image
of \citet*[][]{Dow92}.  Figures \ref{spec1} - \ref{spec2} display
spectra for six nuclear clouds.  Spectra were generated by summing all
the flux within a 6$^{''}$ box centered on the GMC positions (Table
\ref{IntT}).  On each spectrum the expected line position of any other
lines 0.1 K or brighter based on the spectrum of SgrB2
\citep*[][]{T89} are indicated. Line intensities are listed in Table
\ref{IntT}.

One might naively expect that the molecules would follow the basic CO
distribution, since CO is a good overall tracer of molecular gas. Or
perhaps one might expect to find emission peaks for these molecules
preferentially at peaks of HCN peaks, since the molecules of our
sample are high density tracers like HCN. Instead, dramatic variations
in morphology are evident among the different transitions.  Either the
dense gas component of IC 342 has extreme variations in excitation
among the GMCs, or there is widespread chemical differentiation across
the nucleus.  The morphology of the different astrochemical species
provides evidence of changing chemical influences due to star
formation, physical conditions, and dynamics across the nucleus (\S
5).

Fractional abundances ($X$$\rm(mol)\equiv N_{mol}/N_{H_{2}}$) are
listed in Table \ref{AbuT}, based on molecular parameters in Table
\ref{MolP}.  Column densities are determined assuming optically thin
emission, and LTE:
\begin{equation}
N_{mol}~=~ \left(\frac{3k Qe^{E_{u}/kT_{x}}}{8\pi^{3}\nu S_{ul}\mu_{0}^{2} 
g_{K_{u}}g_{I_{u}}} \right)I_{mol},
\end{equation}
where $S_{ul}$, $g$ and $E$ are the line strength, degeneracy and
upper energy of each state, respectively, and $\rm T_{ex}$ is the
excitation temperature associated with the transition.  Given that we
have mapped only one transition of each species, corrections for
background radiation and opacity have been ignored.  Column densities
are sensitive to $\rm T_{ex}$ through the partition function, $Q$, and
the energy of the upper state.  The asymmetric tops (HNCO and
CH$_{3}$OH) are more sensitive to temperature changes than the linear
rotors.  Changes in gas density also affect excitation, particularly
for molecules with high critical densities (HNC, N$_{2}$H$^{+}$ and
HC$_{3}$N).  Fortunately in IC 342 these properties have been at least
partially constrained by observation.  The kinetic temperature, $\rm
T_{k}$, determined from NH$_{3}$ is $\sim$50 K \citep*[][]{HMR82}
which is similar to the derived dust temperature, 42 K
\citep*[][]{Bec80,RH84}.  Modeling of the optically thick lines of
CO(1-0), CO(2-1) and CO(3-2) indicate temperatures of 15-40 K
\citep{HTM87,Eck90,XYS94}; however, the optically thick CO lines may
be biased toward the outer, radiation-warmed layers of the clouds
\citetext{\citealp{THH93};\citeauthor{MT01}}.  We adopt $\rm T_{ex}$ =
10 K determined from C$^{18}$O interferometer maps (\citeauthor{MT01})
and single-dish H$_{2}$CO and CH$_{3}$OH measurements
\citep[][]{HMH97} as the most suitable excitation temperature to use
with the observed tracer species.  Table \ref{MolP} gives the factor
by which the column densities would change if the assumed $\rm T_{ex}$
was changed from 10 K to 50 K.  Given the critical densities of the
mapped species, $\rm T_{ex}$ is very likely lower than $\rm T_{k}$
(subthermal), so 50 K is considered a robust upper limit to the
excitation temperature.

Fractional abundances require, in addition, an $\rm H_2$ column
density, N(H$_{2}$).  N(H$_{2}$) is most easily obtained from the
CO(1-0) brightness and an empirical Galactic conversion factor, $\rm
X_{CO}$.  However, $\rm X_{CO}$ overpredicts N(H$_{2}$) in nearby
galaxy centers, including IC 342, by factors of a few
\citetext{eg. \citeauthor{MT01}; \citealp{DHWM98,WNHK01,MT04}}.  A
better measure of N(H$_{2}$) can be obtained from the C$^{18}$O.  It
is known to be optically thin and for the excitation temperatures
observed, the intensities of the lower J transitions of C$^{18}$O are
not strongly dependent on gas excitation \citetext{see
\citeauthor{MT01} for a detailed discussion of N(H$_{2}$) and its
uncertainties in IC 342}.  Hence we adopt optically thin
C$^{18}$O(1-0) and [H$_{2}$/C$^{18}$O] = $2.9 \times 10^{6}$
([$^{12}$CO/C$^{18}$O] = 250; \citet[][]{HM93} and [CO/H$_{2}$] =
$8.5\times 10^{-5}$; \citet[][]{FLW82}) when calculating the H$_{2}$
column densities, consistent with what is derived from the CO
isotopomers. The lines observed in this study are also optically thin
and have similar upper energy states, and thus their intensities vary
with $\rm T_{ex}$ in step with C$^{18}$O, providing at least partial
compensation for changing physical conditions.  Moreover,
C$^{18}$O(1-0) has a higher critical density ($n_{crit}\simeq 2 \times
10^{3}$ cm$^{-3}$) than the optically thick CO, and should have more
similar beam filling factors to the lines presented here.  If the
highest critical density species are strongly subthermal abundances
may be somewhat overestimated.  On the other hand, for these same high
critical density species, it is expected that their emission will be
more confined than C$^{18}$O and hence underestimated locally.
Together with the compensating effect of C$^{18}$O(1-0) discussed
above, we estimate that the column densities and fractional abundances
are uncertain to at least a factor of three, although the relative
column densities---that is, the relative spatial distributions within
the nucleus---are probably more reliable.

We now introduce each of the different molecular species that have
been mapped and discuss their idiosyncrasies before we turn to the
overall chemical picture in IC 342. Those who are already familar with
the molecules or are easily bored may skip to the next section.

\noindent{\it C$_{2}$H \label{morph_c2h}---Ethynyl:} This is the J =
3/2-1/2 fine structure component of the N = 1-0 transition.  C$_{2}$H
is confined to the central ring, with a peak antenna temperature of
$\rm T_{mb}$ = 0.27 K (Table \ref{IntT}) at GMC A.  Figure
\ref{hr_mol} shows that the C$_{2}$H emission in GMC A does not
originate from the main HCN peak but from its western side.  C$_{2}$H
is also brightest on the inner, starburst lit, face of GMC C, and it
follows the H$\alpha$ wisps (Figure \ref{hst}) between GMCs A and C.
C$_{2}$H emission also appears in the central trough.  The C$_{2}$H
spectrum appears preferentially blueshifted towards the central
regions of IC 342, but this may be an artifact of the presence of the
F = 1-0 hyperfine component.  At GMC A the line is strong enough to
separate the F=1-0 and F=2-1 hyperfine components. Their ratio is 2,
the value expected for optically thin, LTE excitation. The F=1-1
component is not detected.

C$_{2}$H fractional abundances in Galactic cores range from $1
- 60 \times 10^{-10}$ \citep*[][]{WBGLS80,HCK84,W83} and reach $\sim 2
\times 10^{-8}$ \citep*[][]{TTH99,LL00} in Galactic diffuse clouds.
In IC 342, a peak abundance of $3 \times 10^{-8}$,
similar to Galactic diffuse clouds, obtains towards GMC A.  Upper
limits towards the other major GMCs are an order of magnitude lower,
more like Galactic dense cores.

\noindent{\it C$^{34}$S \label{morph_c34s}---Carbon Monosulfide:} The
J = 2-1 rotational transition of C$^{34}$S 
expected to be optically thin.  Like C$_{2}$H, C$^{34}$S(2-1) emission
is confined to the central ring region, and brightest towards GMC A
($\rm T_{mb}$ = 0.16 K). C$^{34}$S, however, lacks the eastern
extension seen in C$_{2}$H.  In general C$^{34}$S(2-1) avoids the
density peaks traced in the \citet*[][]{Dow92} HCN(1-0) image,
although C$^{34}$S also has a high critical density.  As with
C$_{2}$H, the northern extension appears predominately on the inner
face of GMC C, toward the nuclear star cluster.  The C$^{34}$S line at
GMC A is rather broader than seen in the other observed transitions
and appears blueshifted like C$_{2}$H (Figure \ref{spec1}).

For an C$^{32}$S/C$^{34}$S isotopic abundance of 23
\citep*[eg.][however see Chin et al. 1996]{WR94} $X$$(\rm C^{34}S) =
1-3\times 10^{-10}$ in Galactic dense cores
\citep*[eg.][]{Wang93,MELP97,LSJZ98} and diffuse/translucent clouds
\citep*[][]{N84,DKV89,LL02}.  In IC 342, we find $X$$(\rm C^{34}S) = 2
\times 10^{-9}$ towards GMC A.  The upper limits elsewhere are
consistent with the typical Galactic values.

Convolving the C$^{34}$S(2-1) interferometer data to the resolution of
the single-dish map of the main CS isotopomer \citep*[][]{MHWH89},
yields a morphology similar to that seen in the single-dish data
except with a less prominent northern peak.  The
CS(2-1)/C$^{34}$S(2-1) intensity ratio towards the central peak is
$\sim$8, while towards GMC C the intensity ratio increases to
$\gsim$12.  Assuming a Galactic $^{32}$S/$^{34}$S abundance ratio of
23 implies a opacities of $\sim$2 in the main isotope towards GMC A
and slightly lower opacities towards GMC C.  However recent
observations suggest that in starburst nuclei C$^{32}$S/C$^{34}$S
$\sim$ 8 - 13 \citep[][]{WHCWHMM04,MMMHG04}.  In this case,
C$^{34}$S(2-1) has low opacity everywhere in the nucleus.

\noindent{\it HNC \label{morph_hnc}---Hydrogen Isocyanide:} The
J = 1-0 line of the linear molecule HNC is the
brightest of the observed lines with a peak antenna temperature of
$\rm T_{mb}$ = 0.62 K, brighter even than C$^{18}$O(1-0).  HNC peaks at
the starburst GMC B, and is bright at all other labeled GMCs with the
possible exception of GMC E.  The morphology of HNC(1-0) is similar to
that of HCN(1-0) \citep*[][]{Dow92}. HNC and HCN may be the best
tracers of the dense gas distribution (\S4).  HCN emission, and
presumably the dense gas, tends to arise on the clockwise (leading) side
of the molecular arms when compared to $^{12}$CO(1-0), an effect also
seen in CO isotopomers \citetext{\citealp{WITHL93};
\citeauthor{MTH00,MT01}}.  The HNC peak at GMC C is shifted closer to
the nucleus than the $^{12}$CO(1-0).  HNC(1-0) also peaks $\sim
5^{''}$ due east of GMC D, a feature not obvious in the map of
HCN(1-0) \citep*[][]{Dow92}.  This location is bright in several other
lines, particularly N$_{2}$H$^{+}$, CH$_{3}$OH and HNCO.  For the sake
of reference, this location will be referred to as D' and its position
is given in Table \ref{IntT}.

HNC abundances are $5 - 10 \times 10^{-9}$ in Galactic dark
clouds, and about an order of magnitude lower in clouds with massive
star formation, including the Galactic Center
\citep*[][]{WESV78,BSMP87,HYMO98,Num00}, or diffuse and translucent
clouds \citep*[][]{NM89,TPM97,LL01}. In IC 342 HNC abundances are
fairly constant across the galaxy, and consistent with that found in
the Galactic Center.  Uncertain opacity effects could be important,
although we view them as unlikely (\S \ref{quiesgas}).

\noindent{\it N$_{2}$H$^{+}$ \label{morph_n2h}---Diazenylium:} We
observed the ground (J = 1-0) rotational state of this linear
molecule.  The hyperfine splitting is much smaller than the observed
linewidths and is ignored.  Emission from N$_{2}$H$^{+}$ is bright
($\rm T_{mb}$ = 0.21 K) and widespread, nearly matching HNC in extent.
The higher resolution map shows that the N$_{2}$H$^{+}$ emission peaks
between GMCs A and B, even though high resolution images of CO do not
show any local maxima at this location \citep*[][]{SBM03}.  This peak
is also seen in the maps of HC$_{3}$N, HNCO and possibly HNC (Figure
\ref{hr_mol}), and NH$_{3}$ \citep*[][]{HMTJ90}, but no other lines.
This cloud appears only in nitrogen-bearing species, and so we label
it 'N' in Figure \ref{hr_mol}.  On the larger scale, N$_{2}$H$^{+}$ is
similar in morphology to NH$_{3}$, HNCO($4_{04}-3_{03}$), and
CH$_{3}$OH($2_{k}-1_{k}$).  All are bright towards GMC D'.

N$_{2}$H$^{+}$ abundances in Galactic dense cores range are $10^{-10}
- 10^{-9}$, with higher abundances in dark cores
\citep*[][]{WZW92b,BCM98}.  In diffuse and translucent clouds,
N$_{2}$H$^{+}$ has fractional abundances well below $10^{-11}$
\citep*[][]{WZW92b,T95,LL01}.  In IC 342 we find $X$$(\rm N_{2}H^{+})
= 2\times 10^{-10}$ up to $6\times 10^{-10}$ towards GMC D'.  These
abundances are up to an order of magnitude larger than the Galactic
Center values.

\noindent{\it HC$_{3}$N \label{morph_hc3n}---Cyanoacetylene:} We
mapped the J = 10-9 rotational transition of HC$_{3}$N.  This molecule
has the largest electric dipole moment and the highest upper energy
state of the sample (Table \ref{MolP}).  HC$_{3}$N emission is
confined to the two GMCs associated with the youngest starbursts, GMCs
B and C, with peak $\rm T_{mb}$ = 0.20 K at GMC C, where the molecular
arms intersect the central ring.  As with HNC and HCN (Figure
\ref{hr_mol}), the HC$_{3}$N peak towards GMC C is $\sim$ 1/2
beamwidth closer to the nucleus than $^{12}$CO(1-0).  HC$_{3}$N emits
faintly at cloud N.

Abundances of HC$_{3}$N towards Galactic cores are $X$$(\rm HC_{3}N)
\sim few \times 10^{-11} - few \times 10^{-10}$, with the cold cores
towards the high end \citep*[][]{MTPZ76,VLSW83,COM91}.  Translucent
clouds have a similar range \citep*[][]{T98}.  For IC 342, abundances
peak towards GMCs C and D' at $4 \times 10^{-9}$, somewhat higher than
in cold Galactic clouds but much lower than the localized Galactic hot
core values \citep[$>10^{-7}$; eg.][]{DMNC00}.

\noindent{\it HNCO \label{morph_hnco}---Isocyanic Acid:} We observed
the K$_{-1}$ = 0 transition of the J = 4-3 rotational state of the
prolate, slightly asymmetric top, HNCO.  HNCO emission is extended,
with peaks at GMCs C, D' and N, and brightest at D' at $\rm T_{mb}$ =
0.25 K.  HNCO emits only weakly towards the starburst (GMC B) and is
undetected at GMC A.  This transition of HNCO, which has no hyperfine
structure, has the narrowest linewidth of the sample, barely resolved
in the 14 km s$^{-1}$ wide channels.  These line widths, narrow by
extragalactic standards, are typical of Galactic GMCs with massive
star formation, such as SgrB2.  Evidently all galactic rotation has
been resolved out at this spatial resolution and the velocity
dispersion of individual clouds dominate.

Abundances of HNCO range from $X$$(\rm HNCO) \lsim 1\times 10^{-9}$ up
to $3\times 10^{-9}$ towards GMC D'.  On $\sim$1-2 pc scales, Galactic
massive dense cores have abundances of $10^{-9}$ increasing to
$10^{-8}$ as the linewidth of the cloud increases \citep*[][]{ZHM00}.
In translucent clouds abundances are \citep*[$1-3 \times 10^{-9}$;
][]{TTH99}, and up to $>2 \times 10^{-8}$ on sub-parsec sizescales
\citep*[][]{WSCJH96,KS96}.  HNCO abundances in IC 342, averaged over
50 pc scales, are similar to those on $\sim$1 pc scales for massive
cores in the Galaxy.

\noindent{\it CH$_{3}$OH \label{morph_ch3oh}---Methanol:} We observed
the blended set of $2_{1}-1_{1}$E, $2_{0}-1_{0}$E, $2_{0}-1_{0}$A+ and
$2_{-1}-1_{-1}$E low energy, thermal transitions of CH$_{3}$OH
[hereafter designated the $2_{k}-1_{k}$ transition].
CH$_{3}$OH($2_{k}-1_{k}$) emission is as bright as C$^{18}$O(1-0) and
nearly as extensive as $^{13}$CO(1-0)!  The general morphology of
CH$_{3}$OH is similar to C$^{18}$O(1-0) following the $^{12}$CO(1-0)
emission but favoring the leading edges of the nuclear arms.  The
brightest methanol emission comes from GMCs C and D', with peak $\rm
T_{mb}$ = 0.35 K at GMC C.  An additional CH$_{3}$OH peak is seen even
north of GMCs D/D', a position only detected in methanol,
C$^{18}$O(1-0) and HNC(1-0).  CH$_{3}$OH, like HNCO and
N$_{2}$H$^{+}$, is bright towards GMC D' and along the leading edge of
the northern spiral arm.  For a ``hot gas" tracer, methanol is
surprisingly weak near near the starburst at GMC B, actually appearing
as a local minima.  GMC A is also weak in methanol.

Galactic methanol abundances range from $10^{-10} - 10^{-8}$ depending
on source size \citep*[eg.][]{KS94,KDBWA97,MB02}.  Diffuse and dark
cloud CH$_{3}$OH abundances are $2 - 3 \times 10^{-9}$
\citep*[][]{FHMC89,T98}, as determined from thermal transitions.
CH$_{3}$OH abundances in the envelope of SgrB2 are $3 - 10 \times
10^{-9}$ \citep*[][]{Num00}.  On smaller scales, methanol can be
enhanced by factors of a few hundred over dark cloud values in shocks
and outflows \citep*[eg.][]{MWHWSHL86,BP97}.  In IC 342, the observed
range is $3 - 8 \times 10^{-9}$, similar to the envelope of SgrB2.
Multi-line single-dish CH$_{3}$OH data obtain $\rm T_{ex} \simeq$ 5 -
10 K averaged over 200 pc size scales \citep*[][]{HMH97} suggesting
that CH$_{3}$OH emission is not dominated by hot cores.

\noindent{\it Nondetections \label{morph_others}--SO:} We attempted to
image two transitions of SO, the ($2_{3}-1_{2}$) and ($6_{5}-5_{4}$)
transitions.  Upper limits (2$\sigma$) of $\rm I_{SO}<$ 1.5 K km
s$^{-1}$ for the ($2_{3}-1_{2}$) line and $\rm I_{SO}<$ 1.3 K km
s$^{-1}$ for the ($6_{5}-5_{4}$) line are obtained.  The SO abundance
in Galactic dense cores is variable, ranging from $X$$(\rm SO) \sim
5\times 10^{-11}$ up to $1\times 10^{-7}$, with enhancements seen
towards dark clouds and hot cores
\citep*[][]{GGLBP78,RHREKI80,BSMP87}, and in diffuse or translucent
clouds \citep*[$2 - 30 \times 10^{-9}$;][]{T95,HCG95,LL02}.  In IC
342, the upper limits for SO of $5 - 8 \times 10^{-9}$ are not
strongly constraining.

Table \ref{othermolT} presents the limits for fainter transitions
within the bandwidth of the spectrometer.  We note that the broadening
of C$^{34}$S(2-1) could be due to contamination from the
($5_{-2,4}-4_{-2,3}$)E transition of acetaldehyde (CH$_{3}$CHO) at
96.426 GHz.  The \citet*[][]{T89} 3 mm survey finds this transition to
be $\sim$ 1/5 the brightness of C$^{34}$S(2-1) in Sgr B2.  In IC 342,
it would have to be comparable in brightness to C$^{34}$S(2-1) to
explain the line width, which is unlikely.  A $3\sigma$ feature is
also seen in the C$^{18}$O(1-0) bandpass matching the frequency of the
($5_{1,4}-4_{1,3}$)E transition of formamide (NH$_{2}$CHO) at 109.754
GHz.  In the \citet*[][]{T89} survey of Sgr B2 this transition has a
brightness 1/5 of the C$^{18}$O(1-0) line.  Towards GMC E, where the
feature is brightest, the observed ratio is $\sim$1/3.  Therefore, it
is possible that this feature is NH$_{2}$CHO.  CH$_{3}$CHO and
NH$_{2}$CHO have yet to be detected in external galaxies.  These deep
interferometric observations suggest that these two aldehydes may be
worth a dedicated search.

In the rest of the paper, we will omit the transition notations for
the molecules to facilitate the exposition. The reader should be aware
that the correlations that we investigate may be a function of the
energy level and excitation, ie., the specific transition, as well as
the chemistry (\S 5).

\section{Understanding the Molecular Maps: Principal Component Analysis \label{pca}}

The maps give a picture of the astrochemistry of the molecular clouds
in the nuclear region of IC 342. To begin to interpret these maps, we
need to establish similarities and differences between the molecules.
The correlations will reveal trends in what governs the chemistry,
which in turn, can reveal the physical characteristics and forces
within the galaxy that create these conditions.

To quantify the 
morphologies of the molecular maps and star formation, we apply a
principal component analysis (PCA).  PCA is a common technique
\citep*[eg.][]{MH87,K88,ED01,WJ03} used to reduce the dimensionality
of a dataset.  It is useful in identifying a small linear combination
of datapoints that convey a significant percentage of the information
of the whole dataset.  The PCA simplifies the picture of molecular
distribution, reducing a large amount of information to a few images,
providing an excellent framework within which to study the complex
variations in molecular properties in IC 342.

For a description of PCA applied to multi-transition molecular maps,
see \citet*[][]{UBGISS97}.  Each pixel in each map is treated as a
separate datapoint in an $x \times y \times n$ dimensional space,
where $x$ and $y$ are the number of pixels along the corresponding
axis of the maps, and $n$ is the number of maps included.  The
``cloud'' of samples are then projected onto an axis such that the
variance along that axis is a maximum.  This projection corresponds to
the first principal component.  The task is repeated, subject to the
constraint that each successive projection is orthogonal to all
previous projections. Each projection, the principal components or
eigenvectors, will contain decreasing fractions of the total data
variance (assuming there is some correlation in the maps).  As long as
the first few principal components contain a significant fraction of
the variability in the data, the entire dataset can be adequately
described by just these few principal components.

To calculate the principal components for IC 342, the line maps were
convolved to the same geometry and beamsize (6$^{''}$), normalized and
mean-centered.  Eleven maps, $^{12}$CO(1-0), C$^{18}$O(1-0), HCN(1-0)
\citep[][]{Dow92} and 3 mm continuum (\citeauthor{MT01}), plus the
seven detected transitions, were sampled at 1$^{''}$ intervals over
the central $32^{''}\times 62^{''}$, making up a 20824 element
dataspace.  The algorithm used to calculate the principal components
is essentially that of \citet*[][]{MH87}.  The results are displayed
in Tables \ref{PCAcorT} and \ref{PCAT}, and Figures \ref{pcamap} and
\ref{pcavect}.

The correlation matrix resulting from the PCA (Table \ref{PCAcorT})
indicates that all molecules are at least partially correlated. This
is also represented by the most significant correlation, component PC1
(Figure \ref{pcamap}a).  HCN and HNC are the most tightly correlated
of the molecules.  C$^{18}$O, N$_{2}$H$^{+}$, HCN and HNC have large
projections onto PC1, but small projections onto PC2 and PC3 (and the
PC2 and PC3 projections of C$^{18}$O and HNC have opposite signs).
Since HNC is expected to trace the distribution of dense quiescent gas
and C$^{18}$O the column density, it would appear that PC1 represents
the density-weighted average column density map of IC 342.  This also
agrees with the general appearance of PC1. In fact the PC1 map is
basically an average of the C$^{18}$O and HNC maps.  PC1 accounts for
$\sim$2/3 of the variance in the data, and all species have large
projections on PC1.  Variations in density-weighted column density
therefore explain much of the overall morphology of the chemical
species, as one might expect. Molecules are found in molecular clouds.
The molecule N$_{2}$H$^{+}$ projects almost exclusively onto PC1. $\rm
N_2H^+$ in the Galaxy is considered a good ``quiescent" gas tracer,
and that also appears true in the nucleus of IC 342.

HNC, another molecule with a large PC1 projection, is the molecule
most closely correlated with the 3 mm continuum.  Of all of the
detected transitions, HNC has the highest critical density, and is
most heavily weighted toward regions of high density, and not just
high column density. Thus the 3mm continuum is very closely associated
with the dense gas. This extends the findings of \citet{GS04} based on
their studies of global HCN fluxes in galaxies down to GMC sizescales.
However, the correlation of dense gas with star formation, while
excellent, is not perfect; we discuss this in \S 5.

The next principal component, PC2, characterizes the correlations
remaining once density-weighted column density effects are taken into
account. As shown in Figure \ref{pcamap}b, PC2 distinguishes between
molecules that peak at GMC A and those that are extended along the
northern arm, particularly those peaking at GMC D$^\prime$.  HNCO and
CH$_{3}$OH, which are found along the northern arm and are absent in
GMC A, have the largest positive projections on PC2 and are well
correlated.  C$_{2}$H and C$^{34}$S have largest negative projections
onto PC2, appearing almost exclusively at GMC A.  These groups are
anticorrelated in the sense that clouds bright in C$_{2}$H and
C$^{34}$S show little HNCO and CH$_{3}$OH emission and
vice-versa. These differences suggest that these two groups---defined
by the northern arm and GMC A locations--- represent distinct types
of chemistry. We discuss the chemistries of the GMC A vs. northern
arm/GMC D$^\prime$ groups below.  HC$_{3}$N, 3 mm continuum, and
N$_{2}$H$^{+}$ are largely independent of PC2, indicating that their
emission is not strongly influenced by the different chemical
conditions of GMC A vs Northern arm/GMC D$^\prime$.

PC3 is of less significance than the previous two correlations.  PC3
distinguishes between GMC C (particularly between upstream [$^{12}$CO
and C$^{34}$S] and downstream [HC$_{3}$N and 3 mm] species) and
between GMCs C and D.  PC3 thus hints at being connected with the
distribution of massive star formation.


\section{Spatially Resolved Chemistries in the Nucleus of IC 342} 

The large variations in spatial morphology observed for the different
molecules across the central half kiloparsec of IC 342 are caused by
either differences in emissivity (excitation) or abundances
(chemistry).  In this section we discuss the possible influences on
the molecular emission that could lead to the observed variations.

\subsection{Are Differences between the Molecular Species Due to
Excitation?\label{physcon}}

The molecules of our sample have lower opacities, higher critical
densities, and larger partition functions than CO, and so these lines
are more sensitive to changes in excitation due to variations in
density that can affect emissitivity.  Since we already know a fair
amount about cloud conditions in IC 342 from the (1-0) and (2-1) lines
of the CO isotopomers, we can investigate how important excitation is
to the emissivities of these molecules.  All of the detected lines
have similar upper level energies and thus behave similarly to changes
in excitation temperature, $\rm T_{ex}$ (see Table \ref{MolP}).
Significant variations in $\rm T_{ex}$ as traced by C$^{18}$O are not
seen, with a range of temperatures across the nucleus of 7-15 K,
except for a small region towards the starburst (GMC B).  Since the
excitation properties are similar and most of the gas is at a
relatively uniform excitation temperature anyway, $\rm T_{ex}$ is not
the major cause of the spatial differences between the molecules.

Critical density can also play a role in emissivity. The range of
electric dipole moments for the observed species of $\mu_{o}$ = 0.8
(C$_{2}$H) - 3.73 (HC$_{3}$N), corresponds to a factor of $\sim$20 in
critical density ($>$1000 if one counts CO).  The CO isotopomers
indicate that volume densities of the GMCs are $\gsim 10^{4}$
cm$^{-3}$ (\citeauthor{MT01}).  At this density, transitions with the
highest critical density should have the most limited extent.
However, we find that there are extended and confined species at both
lower critical densities (CH$_{3}$OH vs. C$_{2}$H), intermediate
critical densities (HNCO vs.  C$^{34}$S) as well as the highest
critical densities (N$_{2}$H$^{+}$ vs. HC$_{3}$N).  We conclude that
densities in the GMCs are $\sim 10^5$ cm$^{-3}$, high enough to (at
least partially) excite all of the molecules everywhere across the
nucleus, except perhaps HC$_{3}$N and N$_{2}$H$^{+}$, which have the
very highest critical densities.

Although variations in density and temperature are not the primary
forces determining the appearance of the maps of Figure 2, there may
be regions where they play a role.  GMC A is genuinely different from
the other GMCs, in spite of its similarity in CO and HCN.  The first
hints that GMC A was different came from the temperature map made
using C$^{18}$O (\citeauthor{MT01}).  GMC A is a localized `hot spot',
with $\rm T_{ex}~ \simeq$ 15 K, higher than the 10 K we adopt here.
The species with intensities most sensitive to changes in $\rm
T_{ex}$, HNCO and CH$_{3}$OH, would be the most affected by a higher
temperature; the observed weakness of HNCO and CH$_{3}$OH in GMC A
could be due to depopulation of low-lying transitions.  This effect is
also possible for the locized maxima of CH$_{3}$OH towards GMC B.  The
second instance in which excitation appears to play a role is in
HC$_{3}$N, which is closely confined to the starburst sites.  This may
indicate that the sites of the strongest star formation are the
locations with the highest combined density and temperature.

Aside from these exceptions, variations in gas physical conditions do
not appear to determine the widespread morphological changes seen in
the chemical maps.  Since the morphology of the maps are not explained
completely in terms of changes in physical conditions, variation is
the chemistry---relative molecular abundances--- must be important.

\subsection{Do the Maps Reflect Chemical Timescales? 
\label{timedep}}

One difference between molecular clouds of our Galactic disk and the
clouds we observe in the nucleus of IC 342 is timescale. Dynamical
timescales are shorter in galactic nuclei than they are in disks.
Chemical models show that steady-state chemistry obtains after
$10^{6-7}$ yrs, \citep*[eg.,][]{HK73,PH80}, and that abundances of
molecules tend to fall into two categories.  ``Early-time'' molecules,
typically radicals and hydrocarbons --- in general species descending
from C or C$^{+}$ --- are abundant early and get burned into CO and
more complicated species as time passes.  ``Late-time'' molecules,
such as N$_{2}$H$^{+}$, NH$_{3}$ and SO, form from slower
neutral-neutral reactions or are quickly destroyed by abundant C and
C$^{+}$.  Early-time species tend to peak by $\sim 10^{5}$ yrs,
whereas late-time species reach their peak at steady state ($> 10^{6}$
yrs.)  \citep*[eg.,][]{GLF82,W83,MN85}. In the disk of our Galaxy,
these timescales are much shorter than the time between spiral arm
passages, so that steady-state chemistry is expected in the absence of
other disturbances.

In galactic centers, orbital timescales are short enough to rival
chemical timescales.  Based on the rotation curve in the nucleus of
IC~342 \citep*[][]{TH92}, $\tau_{GMC}~\lsim$ 3 Myr at the central
ring, 7 Myr at 20$^{''}$, and 10 Myr at the edge of the molecular
arms.  The arms seen in IC 342 are known to have strong non-circular
and shearing motions, the expected response to a barred potential
\citep{I90,TH92,SBM03}.  It is likely that upon approaching/entering
the arms, molecular clouds are either torn apart by the strong
velocity gradients along the arm or are shocked due to cloud-cloud
collisions, and this will happen for a significant fraction of the
orbit, or every $\sim few \times 10^5~\rm cm^{-3}$.  The ``chemical
clock'' of the molecular clouds could well be reset after traversing
the arms.  Molecular clouds near the dynamical center of the galaxy
may not be able to establish chemical equilibrium between arm
passages; clouds farther out can potentially achieve equilibrium.  A
transition from early-time species to late-time species as
galactocentric radius increases will then be manifested.  

The ``early-time'' species C$_{2}$H and HC$_{3}$N are confined to the
central ring region.  However, other species such as HNC and
CH$_{3}$OH (early-time if produced by gas-phase reactions, e.g. \S
\ref{gasgrain}) are extended.  N$_{2}$H$^{+}$, the most prominent
late-time species detected, is actually brightest towards the central
ring.  Therefore, we conclude that time dependent chemistry is not the
dominant effect causing the chemical differences between the clouds in
IC 342. With the short orbital timescales it is likely that early-time
chemistry is the relevant chemistry over the whole region, although in
this context the brightness of the ``late-time" molecule
N$_{2}$H$^{+}$ is surprising.  In the next section we show that
photo-dissociation region (PDR) chemistry may also create the
appearance of early-time chemistry in this region.

\subsection{The Effects of a High Radiation Field: PDR Chemistry and the Cloud A Peakers C$_{2}$H and C$^{34}$S \label{pdr}}

Photo-dissociation regions are widely believed to be responsible for
many chemical properties of molecular clouds. IC 342 has active
nuclear star formation, with $\rm L_{OB} \sim 10^8~L_\odot$, and a
slightly older ($>$10 Myr) nuclear star cluster.  Clearly PDR
chemistry will be important; however, is it a dominant driver of the
chemistry in IC 342?

The observed molecules most affected by PDR chemistry are C$_{2}$H and
the sulfur-bearing species such as C$^{34}$S. The spatial
distributions of these molecules are distinct from the others (Figure
\ref{pcavect}), and it is likely that C$_{2}$H and C$^{34}$S emission
trace the molecular clouds that are experiencing particularly high
radiation fields.  We discuss the chemistry and detailed distribution
of each below.

\subsubsection{C$_{2}$H} The gas phase chemistry of C$_{2}$H
follows two main pathways.  One is dissociative recombination 
with hydrocarbon ions:
\begin{equation}
C_{2}H_{2}^{+}~~+~~ e^{-}~~\rightarrow ~~ C_{2}H~~ + ~~ H,
\label{c2h_rx1}
\end{equation}
\begin{equation}
C_{2}H_{3}^{+}~~+~~ e^{-}~~\rightarrow ~~ C_{2}H ~~ + ~~ 2H/H_{2},
\label{c2h_rx2}
\end{equation}
where C$_{2}$H$_{2}^{+}$ and C$_{2}$H$_{3}^{+}$ are built up from 
reactions of the form \citep*[eg.,][]{WBGLS80,WWMV88,SD95,THT00}:
\begin{equation}
C^{+}~~+~~ CH_{n}~~\rightarrow ~~ C_{2}H_{2}^{+} ~~ + ~~ H/H_{2},
\label{c2h_rx}.
\end{equation}
The second pathway involves the direct photodissociation of acetylene
(C$_{2}$H$_{2}$), which also forms from reaction (\ref{c2h_rx2})
\cite[eg.][]{TNOOJ87,FMCB93, HJO99}.  C$_{2}$H is destroyed primarily
by photodissociation at A$_{V}$ ($\sim$ 1 mag) and by reactions with O
and C$^{+}$ at A$_{V}$ $\sim$ 5-6 mag
\citep*[eg.,][]{WBGLS80,WWMV88,THT00}.  C$_{2}$H should be abundant
where C$^{+}$ and FUV photons are profuse.  In the Galaxy, C$_{2}$H
abundances of $X_{C_{2}H} \gsim ~ 10^{-8}$ are observed in the diffuse
(A$_{V}~\sim$ 1 - 5m) PDR gas \citep*[eg.][]{FMCB93,HJO99, LL00}.

In IC 342, C$_{2}$H emission is confined to within 40 pc of the
dynamical center, where the radiation field is high due to current
star formation and to the 60 Myr nuclear star cluster.  C$_{2}$H is
bright even in the central molecular ``trough" coincident with the
nuclear star cluster (Figure \ref{hr_mol}). Except in GMC A, C$_{2}$H
emission avoids the density peaks traced by HCN.  In GMCs C and
possibly B C$_{2}$H emission peaks towards the side of the GMC facing
the nuclear cluster. (This could also be true for GMC A, which is
probably viewed ``face-on" and therefore difficult to judge.)
C$_{2}$H appears to arise from cloud surfaces illuminated by the
nuclear star cluster and not by the young IR/radio star-forming
regions. The influence of the IR/radio star-forming regions may be
localized and minimized because these young stars are still deeply
embedded where their photons remain trapped.  GMC A, which is the
brightest in C$_2$H, is the closest cloud to the nuclear cluster.
Apparently photons can penetrate GMC A more effectively than the other
clouds. Perhaps GMC A is more diffuse than the other GMCs, in spite of
its bright HCN emission.

\subsubsection{Sulfur Bearing Molecules: C$^{34}$S} Reactions of S and
S$^{+}$ with hydrogen are endothermic so sulfur chemistry is not
initiated through hydrides.  For CS, formation is primarily via:
\begin{equation}
HCS^{+} ~~~ + ~~~e^{-}~~~ \rightarrow ~~~CS ~~~ + ~~~H, 
\label{rx_hcs}
\end{equation}
where HCS$^{+}$ comes from:
\begin{equation}
S^{+} ~~~ + ~~~ CH/C_{2} ~~~\rightarrow ~~~CS^{+} ~~~ + ~~~H/C, 
\label{rx_s+ch}
\end{equation}
and:
\begin{equation}
C^{+} ~~~ + ~~~SO~~~ \rightarrow ~~~CS^{+} ~~~ + ~~~O, 
\label{rx_c+so}
\end{equation}
followed by reactions with H$_{2}$.  The reaction:
\begin{equation}
C ~~~ + ~~~SO~~~ \rightarrow ~~~CS ~~~ + ~~~O, 
\label{rx_cso}
\end{equation}
can also be important in the formation of CS, particularly when there
is a large SO abundance.  At low A$_{v}$ reaction (\ref{rx_s+ch}) is
should dominate the formation of CS$^{+}$, while at large
A$_{v}$ reaction (\ref{rx_c+so}) dominates.  CS is destroyed by
reactions with atomic O, He$^{+}$ (H$_{3}^{+}$, and HCO$^{+}$ also
destroy CS but the products are rapidly recycled back into CS), and
photodissociation.  Photodissociation dominates the destruction at low
A$_{v}$ \citep*[eg.][]{DKV89,T96}.

SO is considered a `late-time' species built up through the
neutral-neutral reaction,
\begin{equation}
S ~~~ + ~~~OH~~~ \rightarrow ~~~SO ~~~ + ~~~H,
\label{rx_soh}
\end{equation}
and destroyed by reactions (\ref{rx_c+so}) and (\ref{rx_cso}) in
C/C$^{+}$-rich environments.  In environments where carbon has yet to
be locked into CO, SO tends to be burned into CS.  Therefore the
abundance ratio [SO]/[CS]$<<$ 1 at `early-times', in regions of high
C/O elemental abundance and low A$_{v}$ than it is at `late-times',
when [SO]/[CS] $\gsim$10 \citep*[eg.][]{S89,BGSL97,NHBM00}.

C$^{34}$S peaks in the central trough and near GMC A.
The CS abundance may be enhanced in these regions through the increase
in S$^{+}$ associated with a high UV flux from the nuclear cluster.
The CS abundance can be approximated as \citep*[e.g.][]{SD95}:
\begin{equation}
X_{CS} ~\sim ~ \frac{k_{CH}X_{S^{+}}X_{CH} + k_{C_{2}}X_{S^{+}}X_{C_{2}}}
{\left(\frac{G_{o}}{n_{H_{2}}}\right) k_{\gamma}(CS)}.
\end{equation}
If S is undepleted and mostly ionized as expected at modest A$_{v}$,
then for values of $k_{CH}\simeq 6.3\times 10^{-9}$cm$^{3}$ s$^{-1}$,
$k_{C_{2}}\simeq 8.1\times 10^{-10}$cm$^{3}$ s$^{-1}$,
$k_{\gamma}$$\rm \simeq 2\times 10^{-10}exp(-2A_{v})$ s$^{-1}$ and
$[CH]~\sim [C_{2}] \sim 10^{-8}$ \citep*[][]{DKV89}, the observed CS
(23*C$^{34}$S) abundances of few $\times 10^{-8}$ are obtained for
A$_{v}\simeq 3$, when $G_{o}$ = 320 \citep{IB03} and n$_{H_{2}}$ =
$10^{4}$ cm$^{-3}$.  This is consistent with the observed C$^{34}$S
morphology.

The SO/(23*C$^{34}$S) intensity ratio is everywhere less than 0.7 and
less than 0.1 (2$\sigma$) towards GMC A.  While not strongly
constaining, this shows that SO is not enhanced in IC 342's nucleus
and is consistent with S molecule formation in a C/C$^{+}$-rich
environment.

In the PCA analysis, C$^{34}$S is the most anomalously distributed
molecule of all (Figure \ref{pcavect}), with the lowest correlation
coefficients with any other molecule.  We suggest that this may be due
to the abundance distribution of ionized sulfur (S$^{+}$).  In diffuse
PDR cloud edges, sulfur should be relatively undepleted and ionized
\citep*[eg.][]{LDVB88}. The bright C$^{34}$S emission toward GMC A and
toward the edges of GMCs B and C is consistent with this model.  If
so, then the GMC A cloud is relatively well illuminated.  In dense
clouds, sulfur should be depleted onto the grains, and C$^{34}$S
comparatively faint. This is the case for the GMCs other than A and
the edges of B and C towards the nuclear cluster.  CS depletion in the
denser regions of dark clouds is well established in the Galaxy,
though on a much more local scale \citep*[eg.][]{BCLAL01,DHWB02}.

\subsection{Shocks and Gas-Grain Chemistry in IC 342: The Northern
Arm Peakers CH$_{3}$OH,
HNCO}\label{gasgrain}

The pronounced spiral morphology in the nucleus of IC 342 indicates
that shocks are likely to be present, both in the spiral arms and
where the spiral arms meet the nuclear ring. Shocks can influence the
chemistry of molecular clouds by raising gas temperatures, but they
also affect the chemistry by liberating molecules formed through the
processing of grain mantles. Two molecules believed to be produced by
grain processing are CH$_{3}$OH and HNCO.

\subsubsection{CH$_{3}$OH} 
The gas phase chemistry of methanol is simple.  The only formation
mechanism is radiative association of CH$_{3}^{+}$ + H$_{2}$O to form
CH$_{3}$OH$_{2}^{+}$ followed by dissociative recombination
\citep*[][]{MHC91, T98}.  This formation mechanism is too slow to
produce methanol relative abundances greater than $X$(CH$_{3}$OH)
$\sim 1-3 \times 10^{-9}$.  \citep*[][]{LBH96}.  Abundances in
Galactic star forming regions can reach up to $\sim ~ 10^{-7}$
\citep*[eg.][]{MWHWSHL86} and interstellar ice mantles are rich in
methanol $X$(CH$_{3}$OH) $\sim ~ 10^{-6}$) \citep*[][]{STS91}. The
high abundance of the methanol is thus believed to arise from
hydrogenation of CO on grain surfaces.  Methanol is expected to trace
warm ($\rm T_{k}\gsim$ 90 K) molecular gas where mantles have been
evaporated.

In IC 342 methanol emission is widespread and bright and traces out
the leading edges of the two molecular arms.  If methanol were
produced by the evaporation of grain mantles by warm molecular gas,
then $\rm T_{k}~\gsim$ 90 K would be required. Instead, $\rm
T_{k}\lsim$ 50 K (\S 3.2) and $\rm T_{ex}(CH_{3}OH)\sim 10$ K are
indicated \citep{HMH97} for the nuclear region.  Despite of the cool
temperatures, the fractional abundance of CH$_{3}$OH is uniformly at
or above $X$(CH$_{3}$OH) $\sim ~3 \times 10^{-9}$ (Table \ref{AbuT}).
A second method of injecting a large amount of grain mantle material
back into the ISM is through shock disruption of grains.  Grain mantle
material can be liberated either by the localized heating due to the
shock or directly via grain disruption
\citep*[eg.][]{SKARR94,BP97,BNM98}.  In mild shocks ($v_{s} \lsim$ 10
km s$^{-1}$) mantles of grains can be liberated without destroying the
molecules in the process \citep*[eg.,][]{BNM98}.  In order to be
relevant for these observations the shocks must operate coherently
over several tens of pcs in order to generate noticable enhancements
on the observed scales.  Shocks due to local phenomena such as
outflows from massive stars likely do not influence enough of the
molecular gas to explain the elevated abundances unless there are a
very large number of them.

Large scale shocks due to orbital dynamics are the most plausible
explanation for the bright methanol emission.  The central region of
IC 342 must have a barred potential to explain the overall molecular
morphology.  Non-circular orbital motions and highly supersonic
changes in velocity direction are observed, particularly in the
northern arm \citep[eg.][]{TH92}.  Orbital shocks associated with
cloud-cloud collisions would be expected, although the exact nature of
these shocks would depend on the strength and geometry of the magnetic
fields.  We expect shock processing of grains to be most important
along the arms, at the intersections of the molecular arms and the
central ring (the $x_{1} - x_{2}$ orbital intersections: GMCs B and C)
and at the bar ends where the molecular gas piles up and collides with
the existing molecular gas in the arms (GMC D/D').  All of these
locations are sites of bright methanol emission.  While star formation
at GMCs B and C complicates the interpretation at the $x_{1} - x_{2}$
orbital intersections, the enhanced methanol abundance towards GMC
D/D' is almost certainly due to shocks resulting from molecular gas
from the southern arm, ``spraying'' off the central ring and colliding
with the gas already present in the northern arm.

\subsubsection{The Enigmatic HNCO}
The chemistry of HNCO is poorly understood. The distribution in IC 342
and its correlation with other northern arm peakers such as methanol,
give interesting insights on the chemistry of HNCO.

It has been proposed that the excited K ladders (K$_{-1}> 0$) of HNCO
are excited by FIR radiation because their critical densities are
prohibitively high for collisional excitation
\citep*[eg.,][]{CWMM86,WSCJH96,Bet96}.  The situation is less clear
for the K$_{-1}$=0 transitions observed here because the critical
densities are lower.  Rapid $b$-type transitions from K$_{-1}~>$ 0 to
K$_{-1}$ = 0 states can thermalize the level populations at the
330$\mu$m (K$_{-1}$ = 1 - 0) and 110$\mu$m (K$_{-1}$ = 2 - 1)
radiation temperatures, but the lower critical densities make
collisional excitation more relevant.  In both Sgr B2 and OMC-1, the
low lying K$_{-1}$ = 0 transitions have higher column densities and
lower rotational temperatures as compared to the excited K ladders
\citep*[][]{CWMM86,BSMP87}.  Derived abundances of HNCO(K$_{-1}$=0)
can be high, and emission much more widely distributed than found for
the K$_{-1}~\neq$ 0.

In IC 342 HNCO is not well correlated with either molecular column
density traced by $^{13}$CO, dense gas traced by HCN, or massive
star-formation traced by 3 mm continuum.  One would expect that if the
dominant excitation mechanism for HNCO was FIR pumping, it would tend
to be found near regions of active star formation and large gas
column.  That it is not arising from GMCs B and C argues against
excitation mechanisms involving the FIR radiation field.
We conclude that the 3 mm K$_{-1}$ = 0 transitions of HNCO
do not trace the FIR radiation field in IC 342. The HNCO emission 
appears to be governed by abundance variations rather than excitation.

 \citet*[][]{I77} considered the possibility that HNCO is formed from electron
recombination of H$_{2}$NCO$^{+}$, which has been produced by
ion-molecule reactions between NCO$^{+}$ and H$_{2}$.  He found that
this model fell short of the observed abundances by an order of
magnitude. \citet*[][]{TTH99} suggest that the dominant formation
mechanism is through:
$$
CN~~+~~ O_{2}~~\rightarrow ~~ NCO ~~ + ~~O,
$$  
\begin{equation}
NCO ~~ + ~~H_{2}~~ \rightarrow ~~HNCO ~~ + ~~ H,
\end{equation}
and destroyed predominately by reactions with H$_{3}^{+}$ and
He$^{+}$.  For abundances $X$(HNCO) $\lsim~10^{-9}$ typical of the
translucent clouds being studied, they argue that gas phase reactions
alone appear necessary to explain the observed abundances.  While the
first reaction step is rapid, the second possesses a significant
activation barrier ($\sim$1000 K), and therefore the above reaction
scheme may be too slow at typical ISM temperatures.  Additionally,
O$_{2}$ is notoriously underabundant in the Galactic ISM
\citep*[eg.][]{SWAS00,O200}, hence the abundance of NCO is very poorly
known.  So as it stands, gas-phase chemistry alone may able to
generate [HNCO/H$_{2}$]$\sim 10^{-(10-9)}$, but the matter remains
unsettled.

Gas-grain chemical models have no trouble producing large abundances
of HNCO.  Chemical models find that significant amounts of HNCO
($X$(HNCO)$\sim 10^{-6}$) are formed in ice mantles by reactions such
as CO(g) + N(g) + H(g) $\rightarrow$ HNCO and C(g) + N(g)
$\rightarrow$ CN(g); CN(g) + O(g) $\rightarrow$ NCO(g); NCO(g) + H(g)
$\rightarrow$ HNCO \citep*[][]{HH93}.  HNCO can also be formed by
reactions of NH$_{3}$ (and its daughter products) with CO
\citep*[][]{HM00}.  Observers of ice mantles have often ascribed the
so-called 4.62 $\mu m$ 'XCN' feature to the ion, OCN$^{-}$
\citep*[eg.][]{GG87,DDHJHB98,NSK01}, which forms either from UV
photolysis of CO + NH$_{3}$ \citep*[eg.,][]{GG87,SG97} or by acid-base
reactions between HNCO and NH$_{3}$ \citep*[eg.,][] {KTBSW01}.  In
either case, the presence of this feature has been taken as evidence
that HNCO is an abundant constituent of interstellar ices
\citep*[eg.,][]{VB98,VKS04}.  If so, then like SiO and CH$_3$OH it is
reasonable to expect HNCO abundances may also be enhanced in shocked
regions.

Observational evidence is mounting for a shock tracer interpretation
for the production of HNCO.  (1) In SgrB2 the K$_{-1}>$0 transitions
peak locally at SgrB2(N), near warm dust, and likely IR excitation,
whereas the lower critical density K$_{-1}$=0 transitions arise from
the more extended envelope \citep*[eg.,][]{CWMM86,KS96}.  The
K$_{-1}$=0 emission is strong towards the north and west edges of
SgrB2, near the CO `hole' seen in the molecular disk
\citep{KS96,MHHI98}, a hole that may have been created in a
cloud-cloud collision \citep*[eg.,][]{HSWM94,MDGGC95,SHWS98}.  (2)
HNCO is distributed differently from C$^{18}$O and is significantly
enhanced in another Galactic Center cloud known to have strong shocks,
GMC G+1.6-0.025 \citep[][]{MB98,HDMHWM98}. (3) A recent study of dense
cores in the Galactic disk, find a tight correlation between HNCO
(K$_{-1}$=0) and SiO, a well established shock tracer suggesting the
same production mechanism \citep*[][]{ZHM00}.

Towards the northern bar end in IC 342 (GMC D'), $X$(HNCO) can be well
above $10^{-9}$ over $\sim$100 pc scales.  It seems unlikely that
gas-phase chemistry can dominate in this region.  Localized
shocks/mantle evaporation due to star formation such as in the SgrB2
or Orion Hot Core cannot provide the observed enhancement unless star
formation is extreme and widespread here, which it is not.  The PCA
finds that HNCO has a morphology most similar to that of CH$_{3}$OH,
which is expected to trace shocks.  These extragalactic observations
appear to suggest that HNCO is being liberated from grain mantles due
to the action of large scale shocks.

If the shock interpretation for the 3mm HNCO emission is correct, it
is somewhat surprising that the linewidth of HNCO seen towards GMC D'
is quite narrow. However, every other line peaking at GMC D' is
broadened by unresolved hyperfine structure except HNC(1-0), so HNCO's
linewidth may appear ``artifically'' narrow.  The narrow linewidth may
be evidence that the shocks are not strong.  Strong shocks (v$\sim$
100 km s$^{-1}$) are expected to destroy the molecules as they are
released from the grains, and there are some suggestions of this in
the Galaxy \citep*[][]{ZHM00}. Large-scale shocks due to cloud-cloud
collisions at orbital resonances appear optimal in this respect,
because they are probably mild, correlated with magnetic fields, and
operate coherently over many tens of parsecs.

\subsection{An Extragalactic Map of the HCN/HNC Ratio \label{quiesgas}}

The HCN/HNC abundance ratio is also an astrochemical enigma.  In
thermodynamic equilibrium, HCN is highly favored over the rare HNC.
However, in cold dark clouds in the Galaxy, HNC is observed to be of
similar abundance to HCN, and up to a factor of three more abundant
\citep[][]{HYMO98}.  In hotter regions such as the Orion hot core, HNC
is still present but HCN dominates, with [HCN]/[HNC]$\sim$100
\citep*[eg.,][]{GLEKI81,CNW84,SWPRFG92,TPM97}.  This has been
interpreted as a transition from a chemical pathway that forms HNC and
HCN at roughly equal abundances at low temperatures to neutral-neutral
reaction paths with a small activation energy (E$_{a}~ \sim$ 200 K)
that takes over creating HCN at the expense of HNC in warmth
\citep*[][]{SWPRFG92,HYMO98}.  Hence a low HCN/HNC ratio has been
considered a tracer of cool, quiescent gas.

From a theoretical perspective, the HCN/HNC abundance ratio is much
cloudier.  Standard model chemistry predicts a ratio of unity over a
wide range of conditions \citep*[eg.,][]{LBH96}, but with considerable
uncertainties in the reactions.  Recent models have suggested that the
reaction HCNH$^{+}$ + e$^{-}$ $\rightarrow$ HCN/HNC + H either
slightly favors HNC over HCN \citep*[][]{SHNI98} or is equally
probable \citep[][]{TH98}.  If so, in cold dark clouds where cosmic
ray driven ion-molecule chemistry dominates, it may be possible to
explain the low observed HCN/HNC ratios.  However, the dependence of
the ratio on temperature is still largely unexplained.  Given the
observational evidence that the HCN/HNC ratio increases with
temperature, neutral-neutral reactions with small activation energies
that would preferentially destroy HNC in high temperature gas have
been sought.  Two commonly suggested reactions include:
\begin{equation}
 HNC~~~ +~~~ H ~~~ \rightarrow ~~~  HCN ~~~ + ~~~ H, 
\end{equation}
and:
\begin{equation} 
HNC ~~~ + ~~~ O ~~~ \rightarrow ~~~ NH ~~~ + ~~~CO,
\end{equation} 
\citep*[eg.,][]{SWPRFG92}.  The first reaction is especially promising
because it converts HNC directly into HCN, effecting the HCN/HNC ratio
sensitively.  However, theoretical calculations suggest that both
reactions have activation energies too large to contribute at gas
temperatures less than several hundred Kelvin \citep*[][]{TEH96}.  Other
reactions such as:
\begin{equation} 
N ~~~+~~~ CH_{2} ~~~\rightarrow ~~~ HCN ~~~+ ~~~H,
\end{equation} 
have been suggested \citep*[eg.][]{GLEKI81,TPM97},   but
are calculated to be exothermic enough to undergo
isomerization upon relaxation, and therefore actually end up making
equal amounts of HNC and HCN \citep*[][]{HTT00}.

Observations of HCN and HNC in starburst and Seyfert galaxies often
find HCN/HNC $\sim 1$, typical of Galactic dark clouds
\citep*[][]{HWMLDH95,APHC02}.  Moreover, no correlation is seen
between high HCN/HNC ratios and any typical indicator of high gas
temperature.  It is not clear that these observations even support the
notion that HCN/HNC is an indicator of temperature
\citep*[eg.,][]{APHC02}.  However, it is difficult to interpret single
dish ratios given the large physical size they subtend on the galaxy.
It is in this context that we investigate the high resolution
GMC-to-GMC changes in the HCN/HNC ratio in IC 342.

In Figure \ref{hcn/hnc} and Table \ref{RatT}, we use our HNC data
together with HCN data kindly provided by D. Downes \citep*[][]{Dow92}
to present the first high resolution HCN/HNC line ratio map in an
external galaxy.  The HCN/HNC line ratio is $\sim 1-2$, and fairly
constant over the nuclear region. GMC A has the highest ratio of 2.
While GMC A appears somewhat hotter than the ambient gas in IC 342
(\citeauthor{MT01}), it is not the hottest GMC (GMC B, $\rm
T_{ex} \sim$ 20 K) nor is it near the present star formation and IR
sources (GMCs B and C).  In fact, starburst GMCs B and C are actually
local minima of HCN/HNC.  Finally, though the interpretation is
limited somewhat by the smaller primary beam of the Plateau de Bure
HCN(1-0) data, it does appear that there is a real trend for the
HNC/HNC line ratio to decrease below unity as one goes to large
distances from the center along the northern and southern spiral arms.

If we assume that HCN/HNC traces temperature and we use the
\citet*[][]{HYMO98} empirical Galactic dependence, then across the
entire nuclear region this ratio would predict that $\rm T_{k}$ is
approximately constant at 22 - 27 K.  This is low compared to the
observed dust temperature, 42 K, and gas kinetic temperatures of
$\sim$50 K in ammonia or the observed antenna temperature of the
$^{12}$CO(2-1) transition \citep*[][]{THH93,SBM03}.  The explanation
for the low HCN/HNC values could be that IC 342 has an abundant dense
component that is significantly cooler and more uniform than the more
diffuse component traced by CO. We consider this unlikely, since the
CO, HNC, and HCN distributions are nearly coextensive.  We consider a
more likely the possibility that HCN/HNC is simply not a function of
temperature.

We have assumed that the HCN(1-0)/HNC(1-0) line ratio reflects the
abundance ratio. This is only true if the line opacities are small.
For $\tau > 1$, HCN(1-0)/HNC(1-0)$\sim$1, independent
of chemistry.  Such an effect is most relevant where the line
intensities are the highest, namely GMCs B and C.  This is where the
ratio approaches unity despite the proximity to HII regions.  However
single-dish observations of H$^{13}$CN(1-0) in IC 342 show that
HCN(1-0)/H$^{13}$CN(1-0) $\simeq$ 30 \citep*[][]{SGKK01}, close to
what is expected for the $^{12}$C/$^{13}$C isotopic ratio in the
nuclei of starburst galaxies \citep*[eg.,][]{HM93,WR94}, indicating
low opacities.  Future high resolution observations of H$^{13}$CN and
HN$^{13}$C should provide information on this issue.

\subsection{Quiescent Gas Tracer? N$_{2}$H$^{+}$} 

It has been suggested that N$_{2}$H$^{+}$ is a robust tracer of the
dense, quiescent component of molecular gas \citep*[eg.,][]{WZW92b}.
N$_{2}$H$^{+}$ abundances tend to be low in hot-cores, outflows and
PDRs in the Galaxy. Hence, the extent and brightness of N$_{2}$H$^{+}$
in the nucleus of IC 342, with its active star formation, radiation,
and shocks, is surprising.  One might infer that dense gas is present
in IC 342, but that this gas is largely sterile and quiescent with
``dark cloud-like'' chemical conditions.  In this picture, shocks,
winds and outflows associated with star formation are confined to
small regions around each forming star, and the remaining dense gas is
largely undisturbed.  This would have interesting ramifications for
the extremely luminous ($\rm L_{IR}\sim 10^8~L_\odot$) star forming
GMCs, B and C, which are both strong sources of N$_{2}$H$^{+}$
emission.

Alternatively, there could be differences in the formation or
destruction rates of N$_{2}$H$^{+}$.  N$_{2}$H$^{+}$ is formed almost
exclusively from the reaction:
\begin{equation}
N_{2}~~+~~H_{3}^{+} ~~\rightarrow ~~ N_{2}H^{+} ~~ + ~~H_{2},
\label{n2hform}
\end{equation}
and is destroyed primarily by dissociative recombination with
electrons in diffuse gas, and the ion-neutral reactions with CO and O
in dark clouds.  Therefore, N$_{2}$ is cycled through N$_{2}$H$^{+}$,
and if the abundance of H$_{3}^{+}$ and electrons can be assessed then
N$_{2}$H$^{+}$ is a direct tracer of the (otherwise invisible)
molecular nitrogen \citep*[eg.][]{WZW92a,T95,BCM98}.

The H$_{3}^{+}$ abundance is formed from cosmic-ray ionization of
H$_{2}$ and destroyed by reactions with e$^{-}$, CO and O.  In
steady-state:
\begin{equation}
X(H_{3}^{+}) ~=~ \frac{\zeta/n_{H_{2}}}
{k_{H3,O}X(O) + k_{H3,CO}X(CO) + k_{H3,e}X(e^{-})},
\label{h3eq}
\end{equation}
where $\zeta$ is cosmic-ray ionization rate and the rate coefficients,
$k$, are taken from \citet*[][]{BCM98} and \citet*[][]{DL84}, except
for the dissociative recombination rate ($k_{H3,e}$), which is taken
from \citet*[][]{Mnat03}.  If we assume that N$_{2}$H$^{+}$ is formed by
(\ref{n2hform}) then:
\begin{equation}
\frac{X({N_{2})}}{X(N_{2}H^{+})} ~=~ \frac{k_{N2H,CO}X(CO) 
+ k_{N2H,e}X(e^{-})}{k_{H3,N2}X(N_{2})X(H_{3}^{+})},
\label{n2_n2h}
\end{equation}
or:
\begin{equation}
\frac{X({N_{2})}}{X(N_{2}H^{+})} ~=~ \frac{[k_{N2H,CO}X(CO) 
+ k_{N2H,e}X(e^{-})][k_{H3,O}X(O) + k_{H3,CO}X(CO) + k_{H3,e}X(e^{-})]}
{k_{H3,N2}X(N_{2})[\zeta /n_{H_{2}}]}.
\end{equation}

In this steady-state chemical scheme, the abundance of N$_{2}$H$^{+}$
increases with (1) increasing $\zeta$, (2) decreasing density, (3)
decreasing electron fraction, $X(e^{-})$, or (4) increasing N$_{2}$,
assuming that the CO and O abundances remain fairly constant across
the nucleus.  It seems unlikely that explanations (2) and (3) dominate
in the nucleus of IC 342.  We know the densities are high, at least a
few $\times 10^{4}$ cm$^{-3}$, and $X(e^{-})$ could be as high as
$\simeq ~ 1.5\times 10^{-5}$ based on the C$^{+}$ abundance determined
by \citet*[][]{CGTW85} and \citet*[][]{Eck90}. The most likely
alternatives for the bright, extensive $\rm N_2H^+$ emission are
therefore either a high cosmic-ray ionization rate or enhanced N$_{2}$
abundance.

Figure \ref{Fn2_n2h} displays the $X(N_{2})/X(N_{2}H^{+})$ ratio as a
function of electron fraction, for different values of
$\zeta/n_{H_{2}}$ (see eq.[\ref{n2_n2h}]).  It has been assumed that
$X(O)~\simeq ~ X(CO)=10^{-4}$.  The maximum N$_{2}$ abundance possible
is one half the total cosmic N abundance, or $X(N_{2})~ \sim ~ 4\times
10^{-5}$ \citep*[eg.,][]{AG89}.  The brightness of the high density
tracers suggest that $n_{H_{2}}$ is at least $10^{4-4.5}$ cm$^{-3}$,
implying that $\zeta_{IC\,342} ~> ~ 10^{-17}$ s$^{-1}$, at least the
Galactic disk value.  If some N is not in N$_{2}$H$^{+}$ \citep[and
there is bright ammonia emission][]{HMTJ90} or $n_{H_{2}}$ is greater
than $10^{4}$ cm$^{-3}$, then $\zeta$ is constrained to be higher than
the Galactic rate.  For $X$$\rm (e^{-})\gsim 10^{-7}$, the cosmic ray rate
would have to be greater still. Therefore, to explain the bright and
ubiquitous N$_{2}$H$^{+}$ emission it is likely that $\zeta ~> ~
10^{-17}$ s$^{-1}$ in the nucleus of IC~342.

\subsection{The Warm and the Dense: HC$_{3}$N} 

HC$_{3}$N has the highest upper energy state and one of the highest
critical densities of our sample.  The most striking feature of the
HC$_{3}$N(10-9) intensity distribution is how closely it follows the 3
mm continuum emission.  We consider two possibilities for the bright
HC$_{3}$N emission in regions where the millimeter continuum is high,
(1) enhanced HC$_{3}$N abundance towards star-forming GMCs B and C
(and possibly E) or (2) HC$_{3}$N is more highly excited by collisions
with molecular hydrogen at the sites of the densest and warmest gas.

HC$_{3}$N formation is uncertain, but probably forms from the
neutral-neutral reaction,
\begin{equation}
C_{2}H_{2}~~+~~ CN~~\rightarrow ~~HC_{3}N ~~ + ~~H,
\label{hc3n_rx}
\end{equation} 
and is destroyed by photodissociation and reactions with C$^{+}$
\citep*[eg.][]{SIMS93,FO97,TLH98}.  Since the central ring region has
high abundances of C$_{2}$H it is reasonable to expect this region
will also have high abundances of CN and C$_{2}$H$_{2}$, species that
can maintain large abundances in PDRs.  It is possible, therefore,
that the high estimated HC$_{3}$N abundances follow from CN and
C$_{2}$H$_{2}$ abundances here.  The weakness of HC$_{3}$N towards GMC
A would then be attributed to increased destruction rates due to
photodissociation and large C$^{+}$ abundances and more diffuse gas
working in tandem.

The second possibility for the HC$_{3}$N morphology is excitation.
HC$_{3}$N is the molecule most sensitive to decreases in $\rm T_{ex}$
(decreases in density; Table \ref{MolP}).  If there is an excitation
gradient decreasing with distance from the starburst the abundances
derived outside the central starburst region would be artificially
low.  Given the sensitivity of the derived HC$_{3}$N abundance to
excitation, we do not at this stage overinterpret possible changes in
HC$_{3}$N chemistry and conclude that decreasing excitation away from
the starbursts is the simplest explanation of the HC$_{3}$N
morphology.

If excitation is the answer then why isn't HC$_{3}$N also bright at
GMC A, since N$_{2}$H$^{+}$, HNC (and HCN) are bright there, species
with critical densities nearly as high or higher than HC$_{3}$N? GMC A
is actually one of the warmer GMCs (in CO), so chemical changes may be
favored.  However there is another possibility.  Electrons could be
responsible for some of the collisional excitation towards GMC A.  In
Table \ref{MolP} the critical densities for electron collisions are
compared with those for H$_{2}$ collisions.  If $X$(e$^{-}$) =
$X$$(\rm C^{+})$ $\simeq~1.5\times 10^{-5}$ in the center of IC 342
\citep*[][]{CGTW85,Eck90}, then electrons are of equal or greater
importance as a collision partner for N$_{2}$H$^{+}$, HCN and HNC.
For all other species, with the possible exception of C$^{34}$S(2-1),
collisions with electrons are not relevant.  The C$^{+}$ abundance is
almost certainly lower than the above values everywhere except GMC A
and the central trough given that the C$^{+}$ value was made with a
large beam and likely is dominated by the more diffuse molecular cloud
component assoicated with the PDRs.  Evidently, in the regions with
abundant C$^{+}$, HNC and HCN can be efficiently excited by electrons
\citep*[][]{DPGPR77,TPM97}.  This may have important consequences for
densities derived from HCN in starburst galaxies dominated by PDR
regions.

\subsection{The Overall Chemistry of IC 342 \label{overall}}

When we consider all the observed molecular lines together a
consistent picture of the structure of the ISM in IC 342 begins to
emerge, which is summarized in the cartoon of Figure \ref{schem}.  Gas
response to a barred potential has set up the molecular minispiral
within the nucleus of IC 342.  Energy dissipation from cloud
collisions and star formation within the molecular arms, and angular
momentum transfer from tidal torquing result in a gradual inward drift
of molecular gas along the arms. Molecular gas piles up at the
intersections of the arms and the central ring, where it triggers star
formation at the rate of $\sim 0.1~M_\odot~yr^{-1}$.  In turn, the bar
sets up density gradients both azimuthally along the molecular arms
and radially. The molecular gas densities tend to be higher on the
leading (counter-clockwise) edges of the molecular arms, becoming more
diffuse behind the arms \citetext{\citealp{WITHL93};
\citeauthor{MT01}}. Densities also tend to be higher as the gas
approaches the nucleus, peaking near the $x_{1}$ - $x_{2}$
intersection. Compare the morphology of $^{12}$CO(1-0),
C$^{18}$O(1-0), HNC and HC$_{3}$N in Figure \ref{IntI} to see the
density progression from diffuse to dense gas.

The sphere of influence of the nuclear bar extends not just to the
physical conditions of the gas (density, temperature) but also to the
chemistry.  CH$_{3}$OH and HNCO delineate the locations of shocks
associated with cloud-cloud collisions within spiral arms, following
the theoretical expectations of the locations of shocks in a barred
potential \citep*[eg.][]{A92}.  These molecules indicate that grain
mantle liberation is important towards GMC D' and possibly GMC C along
the northern arm. The lack of the expected symmetry between the
northern and southern arms appears to be due to a lower column density
of dense gas towards the southwestern arm (Figure \ref{pcamap}a).
Outside of GMC A it may be that grain liberation influences the
abundances of most of the observed species either directly (CH$_{3}$OH
and HNCO) or indirectly due to injection of ``invisible'' N$_{2}$ and
C$_{2}$H$_{2}$ (N$_{2}$H$^{+}$ and HC$_{3}$N).  It is tempting to
speculate that the prevalence of molecular arm shocks explain why star
formation is not pronounced along the spiral arms, but ``waits'' till
the gas arrives in the central ring.

From C$_{2}$H and C$^{34}$S, we learn that the faces of the molecular
ring surrounding the central star forming complex are bright PDRs.
The limited spatial extent of C$_{2}$H argues that the C$^{+}$
abundance is strongly enhanced in the central ring; the primary
ionization sources are not distributed uniformly across the nucleus,
nor does the ionization appear to penetrate deeply into most of the
GMCs.  This rules out more penetrative mechanisms such as cosmic-ray
ionization (primary or secondary) or hard X-rays as the main source of
enhanced C$^{+}$ ionization in IC 342.  Photoionization from stars is
likely to produce enhanced C$^{+}$ given the interstellar radiation
field (ISRF) is $\sim$320 times stronger than Galactic value
\citep*[][]{IB03}.  This provides further support for previous studies
which suggested that PDRs play an important role in the physical
conditions of the central ring of IC~342 \citetext{\citealp{THH93};
\citeauthor{MTH00}; \citealp{SGKK01}}.  Synchrotron emission
\citep[Figure \ref{hr_mol}a;][]{TH83} indicates that supernovae have
taken place in the nucleus, and one might expect the cosmic-ray
ionization rate to be somewhat elevated here.  The bright
N$_{2}$H$^{+}$ emission alludes to such a possibility. Since the PCA
analysis demonstrates that N$_{2}$H$^{+}$ and C$^{18}$O have similar
morphologies, whatever process elevates the N$_{2}$H$^{+}$ abundance
operates fairly uniformly across the entire region, as would be
expected for cosmic-ray ionization.

From the morphology of C$_{2}$H, the most likely source of ionizing
radiation for the PDRs (see especially the inner face of GMC C) is the
more mature ($\sim$ 60 Myr old) central, nuclear cluster and not the
youngest ($\sim$ few Myr) starbursts seen in radio and millimeter
continuum towards GMCs B and C.  Evidently, the young stars in these
large HII regions have yet to break out of their natal cocoons and
influence the large-scale chemistry of the nucleus.  The HST
composite image, the preferentially blueshifted C$_{2}$H and C$^{34}$S
line profiles and the weak C$_{2}$H extention perpendicular to the
major axis imply that this central star formation event is just
beginning to break out along the galactic minor axis.

GMC A appears to be the molecular cloud most strongly penetrated by UV
photons.  GMC A is also the only GMC in the central ring that is not a
site of strong current star formation.  These results suggest that for
some reason GMC A is on average more diffuse than the other two GMCs,
and therefore more susceptable to UV radiation.  The reason for this
is unclear, but may be related to its location in the nucleus.  Close
to the dynamical center of the galaxy and at a position where it
likely has just `whipped' around from the $x_{1}$ orbital onto the
$x_{2}$ orbital, tidal and shearing forces may be shredding Cloud A
apart (note the broad C$^{34}$S line profile).  In this condition,
Cloud A is probably not inclined to collapse and form stars (until 
it collides with GMC B+E?).

It has been established that the chemistry of the two nearest
prototypical starburst galaxies, M 82 and NGC 253, have different
chemical makeups \citep*[eg.][]{MH91a,HBM91,MH93}.  M 82 is bright in
HCN, HNC, C$_{2}$H, CN, HCO$^{+}$ and HCO, but notoriously weak in
lines of SiO, CH$_{3}$OH, CH$_{3}$CN, HNCO and SO, as compared to NGC
253 \citep{MH93,TNK95}.  In IC 342, we see for the first time regions
with M 82-like chemistry (GMC A), and regions of NGC 253-like
chemistry (GMC D') in the same nucleus.  Regions where grain
disruption due to shocks occur are likely to manifest chemistries
similar to NGC 253, whereas sites of PDRs manifest M 82-like
chemistries.  It is therefore tempting to conclude that the ISM in M
82 is dominated by GMC A-type molecular gas (PDRs) and is deficient in
GMC D-type molecular gas (shock/grain process species) due to absence
of strong nucleus-wide shocks compared to the barred galaxy, NGC 253.
This lends support to similar conclusions reached by
\citet*[][]{GMFN00} and \citet*[][]{GMFUN02} for NGC 253 and M 82.  It
also illustrates that CH$_{3}$OH and HNCO hold the potential of being
an much more useful shock tracers in external galaxies than the well
established tracer, SiO, since the 3 mm transitions of these two
molecules are much brighter than SiO \citep*[][]{NHJM91,SZ95,HMH97}.  In IC
342 PDRs still remain largely confined to the central ring, but in M
82 they are clearly much more pervasive.

Finally, an intriguing result of this dataset is the apparent
discovery of an isolated nitrogen-rich GMC.  Its location between GMCs
A and B suggests two possibilities.  Firstly, it is seen at the
location where gas going down the northern arm collides with the
southern arm.  This may then represent a location of N$_{2}$ injection
(from grain disruption), which would explain the increased
N$_{2}$H$^{+}$ abundance and HNCO, but not necessarily the HC$_{3}$N
abundance (unless C$_{2}$H$_{2}$ is also injected).  A second
possibility is that this region is enriched in ejecta from young,
massive stars, perhaps Wolf-Rayet stars.  There is a young, massive
cluster seen in the near-infrared (NIR) near this location \citep*[see
Figure 3 of][]{SBM03}.  There is evidence that massive star enrichment
of $\rm ^{18}O$ has taken place in the central trough vicinity
(\citeauthor{MT01}).  But why is the nitrogen enrichment seen in this
one location when there are several other NIR bright clusters,
including the much larger central cluster, which is not known to be
enriched?  Similar studies of other galaxies may shed light on this
question.

\section{Conclusions \label{conc}}

We have detected emission from a number of molecules in the central
region of IC 342 at resolution of 5\arcsec\ using OVRO.  Significant
differences are seen in the spatial distributions of these molecules,
and we have analyzed the causes of these differences.  Main results 
include:

1. A principal component analysis of the maps of the seven detected
molecules, $^{12}$CO(1-0), $\rm C^{18}O$, HCN(1-0) and 3mm continuum,
reveals that the maps all have some degree of correlation, and that
the dominant common factor (PC1 axis) is the density-weighted mean
column density. Molecules with the largest projections on PC1
are $\rm N_2H^+$, $\rm C^{18}O$, and HNC. These molecules, and
primarily $\rm N_2H^+$, give the best overall representation of the
total molecular gas distribution.

2. Differences in excitation and critical density are not the dominant
influences shaping the spatial differences among the 
different molecules in IC342, since the excitation energies of the
transitions are similar, and cloud densities are high enough that most
molecules are collisionally excited.

3. The spatial differences among the molecular line emission are not
driven by variations in chemical timescale: all clouds in this region
are likely to have early-time chemistry due to the short orbital
timescales of the nucleus.

4. The second principal component axis, PC2, splits into two
anti-correlated groups of molecules: molecules that peak within the
central 75 pc  near GMC A and the nuclear star cluster, and
those that peak up along the prominent northeastern arm, located
100-200 pc away from the nucleus.

5. The GMC A peakers, $\rm C_2H$ and C$^{34}$S, are enhanced in the
center of IC 342 by high radiation fields. GMC A is a PDR cloud. The
fact that these molecules tend to favor the parts of the clouds
towards the center of IC 342 rather than toward the slightly
off-center $\rm L_{IR}\sim10^8~L_\odot$ radio/IR source suggests that
the source of the radiation field affecting the chemistry of GMC A is
the $\sim$60 Myr central star cluster and not the actively star
forming radio/IR sources.

6. Molecules that peak along the north-east arm of the molecular
mini-spiral are HNCO and $\rm CH_3OH$.  Large scale shocks due to
the changes in gas velocity at the arms of the nuclear minispiral are
probably the cause of enhanced HNCO and $\rm CH_3OH$ here.

7. HCN and HNC are the best tracers of the dense gas and HC$_{3}$N of
warm, dense gas. Their correlation with the 3mm continuum is
excellent. The HCN/HNC ratio is 1-2 across the nuclear region, and
appears to be unrelated to kinetic temperature.

8. The brightness and extent of emission from $N_2H^+$ suggests that
the cosmic ray ionization rate is higher than $\zeta ~> ~ 10^{-17}$
s$^{-1}$ in the nucleus of IC~342.

\acknowledgements 
 
We appreciate the support and assistance of the faculty, staff and
postdocs at OVRO during the observations.  We are grateful to
D. Downes for kindly providing us access to the HCN(1-0) data.  We
thank M. Jura and L. E. Snyder for reading drafts of the paper, and
E. C. Sutton, M. Morris, and R. Hurt for helpful discussions.  An
anonymous referee is also thanked for providing a detailed, helpful
and prompt report. DSM acknowledges support from the Laboratory for
Astronomical Imaging at the University of Illinois through the NSF
grant AST-0228953.  This work is also supported by NSF grants
AST-0071276 and AST-03079950 to JLT.  The Owens Valley Millimeter
Interferometer is operated by Caltech with support from the NSF under
Grant AST-9981546.

\clearpage

\begin{deluxetable}{lccccccc} 
\rotate
\footnotesize
\tablenum{1} 
\tablewidth{0pt} 
\tablecaption{Observational Data\label{ObsT}} 
\tablehead{ 
\colhead{Transition}  
&\colhead{Dates}
&\colhead{Frequency}
&\colhead{T$_{sys}$}
&\colhead{$\Delta V_{chan}$} 
&\colhead{$\Delta \nu_{band}$}  
& \colhead{Beam}  
& \colhead{Noise} \\ 
\colhead{}  
&\colhead{\it (MMYY)}
&\colhead{\it (GHz)}  
&\colhead{\it (K)}
&\colhead{($km~s^{-1}$)} 
&\colhead{\it (MHz)} 
& \colhead{\it ($^{''}\times^{''}$;$^{o}$)}  
& \colhead{\it (mK/mJy bm$^{-1}$)}}  
\startdata 
C$_{2}$H($1-0;3/2-1/2$)\tablenotemark{a}& 0900-1000& 87.317 &350-690 
& 13.73& 128 &$5.5\times 4.9$;-41$^{o}$  & 53/8.9 \\ 
HNCO($4_{04}-3_{03}$)\tablenotemark{a}& 0900-1000& 87.925 &420-780 
& 13.64& 128 &$5.9\times 5.1$;-36$^{o}$  & 39/7.4 \\ 
HNC($1-0$)\tablenotemark{a}& 0900-1000& 90.664 &430-930 
& 13.23& 128 &$5.9\times 5.1$;-39$^{o}$  & 26/5.2 \\ 
HC$_{3}$N($10-9$)\tablenotemark{a}& 0900-1000& 90.979 &400-890 
& 13.18& 128 &$5.9\times 5.1$;-44$^{o}$  & 24/4.8 \\ 
N$_{2}$H$^{+}$($1-0$)\tablenotemark{a}& 0101-0401& 93.174 &380-440 
& 12.87& 128 &$6.2\times 5.1$;-59$^{o}$  & 29/6.4 \\ 
C$^{34}$S($2-1$)\tablenotemark{a}& 0101-0401& 96.413 &370-450 
& 12.44& 128 &$6.6\times 5.6$;-63$^{o}$  & 24/6.8 \\ 
CH$_{3}$OH($2_{k}-1_{k}$)\tablenotemark{a}& 0101-0401& 96.741 &420-480 
& 12.40& 240 &$6.0\times 4.8$;-65$^{o}$  & 35/7.6 \\
SO($2_{3}-1_{2}$)\tablenotemark{b}& 1097-1297& 109.252 &410-480 
& 10.98& 128 &$5.0\times 4.3$;-21$^{o}$\tablenotemark{c}  & 41/8.6 \\
SO($6_{5}-5_{4}$)\tablenotemark{b}& 1097-1297& 219.949 &410-1020 
& 5.45& 128 &$5.0\times 4.3$;-21$^{o}$\tablenotemark{d}  & 52/44 \\
\enddata 
\tablenotetext{a}{Phase Center \#1: $\alpha = 03^{h} 41^{m} 57^{s}.0~~ 
\delta = +67^{o} 56' 30.^{''}0$ (B1950); v$_{lsr}$ = 35 km s$^{-1}$} 
\tablenotetext{b}{Phase Center \#1: $\alpha = 03^{h} 41^{m} 57^{s}.0~~ 
\delta = +67^{o} 56' 26.^{''}0$ (B1950); v$_{lsr}$ = 35 km s$^{-1}$ \\ 
$~~~~$Phase Center \#2: $\alpha = 03^{h} 41^{m} 57^{s}.9~~ 
\delta = +67^{o} 56' 29.^{''}0$ (B1950); v$_{lsr}$ = 35 km s$^{-1}$} 
\tablenotetext{c}{Maps generated with a 50 k$\lambda$ taper.}
\tablenotetext{d}{Map convolved to the same resolution as the 3 mm 
SO transition.}
\end{deluxetable} 
 
\clearpage
 
\begin{deluxetable}{lcccccccc} 
\rotate
\footnotesize
\tablenum{2} 
\tablewidth{0pt} 
\tablecaption{Molecule and Column Density Parameters\label{MolP}} 
\tablehead{ 
\colhead{Molecular}  
&\colhead{$\mu$}
&\colhead{A/B/C}
&\colhead{$S_{ul}g_{K}g_{I}$ }
&\colhead{$E_{u}$}    
&\colhead{$^{H_{2}}n_{cr}$/$^{e}n_{cr}$\tablenotemark{a} } 
&\colhead{$\frac{N(50)}{N(10)}$\tablenotemark{b}}  
&\colhead{T$_{min}$\tablenotemark{c} } 
&\colhead{$\frac{N_{min}}{N_{10}}$\tablenotemark{d}}  
\\
\colhead{Transition}  
&\colhead{\it (Dby)}
&\colhead{\it (GHz)}  
&\colhead{}
&\colhead{\it (K)}  
&\colhead{\it log(cm$^{-3}$)} 
&\colhead{} 
&\colhead{\it (K)} 
&\colhead{} 
}
\startdata 
C$_{2}$H:($1-0;\frac{3}{2}-\frac{1}{2}$)& 0.8&\nodata/43.675/\nodata 
&1 &4.19 &5.13/6.06 &3.01 &4.80&0.66 \\
HNCO($4_{04}-3_{03})$&1.60 &912.711/11.071/10.911 &4 &10.55 &5.43/6.03 &4.81 &7.03&0.92  \\
HNC($1-0$)& 3.05&\nodata/45.332/\nodata &1 & 4.35 & 6.39/5.69&2.94 &4.98&0.70  \\
HC$_{3}$N($10-9$)&3.72 &\nodata/4.5491/\nodata &10 &24.02 &5.86/6.69 &0.723 &24.0&0.59 \\
N$_{2}$H$^{+}$($1-0$)&3.40 &\nodata/46.587/\nodata &1 &4.47 &5.63/5.67 &3.00 &5.12&0.75 \\
C$^{34}$S($2-1$)&1.96 &\nodata/24.104/\nodata &2 &6.94&5.65/5.99 &2.56 &7.29&0.61 \\
CH$_{3}$OH($2_{k}-1_{k}$)&0.89 &127.484/24.680/23.770 &4,3,4\tablenotemark{e} 
&6.98 &4.83/6.21 &5.96 &4.65&0.65 \\
SO($2_{3}-1_{2}$)&1.55 &\nodata/21.523/\nodata &1.51&21.06 &5.27/6.10&0.901 &21.4&0.68  \\
SO($6_{5}-5_{4}$)&1.55 &\nodata/21.523/\nodata &5.91&34.99&6.55/7.30 &0.296 &35.3&0.28 \\
\enddata 

\tablecomments{Data from the JPL Molecular Spectroscopy Catalog
\citep[][]{JPLmol} and references within.  Collisional coefficients
are HNC, C$_{2}$H \citep[HCN;][]{GT74}, C$^{34}$S, HC$_{3}$N
\citep[][]{GC78}, N$_{2}$H$^{+}$ \citep[][]{G75}, CH$_{3}$OH
\citep[][]{PFD04}, HNCO \citep[][]{G86} and SO \citep[][]{G94}.}

\tablenotetext{a}{The critical density of the transitions neglecting
opacity effects ($^{H_{2}}n_{cr}\simeq \frac{A_{ul}}{C_{ul}}$) and
$^{e}n_{cr}$ based on the formalism of \citep*[][]{DPGPR77}, with 
[e$^{-}$] = [C$^{+}$] $\simeq ~ 1.5 \times 10^{-5}$.}
\tablenotetext{b}{The ratio by which the derived column densities 
change when the $\rm T_{ex}$ is changed from 10 K $\rightarrow$ 
50 K.}
\tablenotetext{c}{The $\rm T_{ex}$ at which derived (LTE) column densities 
are a minimum.}
\tablenotetext{d}{The ratio by which the derived column densities 
change when the $\rm T_{ex}$  is changed from 10 K $\rightarrow$ 
T$_{min}$.}
\tablenotetext{e}{$S_{ul}g_{K}g_{I}$ for the three blended $2_{1}-1_{1}$E, 
$2_{0}-1_{0}$E, and $2_{0}-1_{0}$A+ transitions \citep[eg.,][]{T91}.}
\end{deluxetable} 
 
\clearpage 
 
\begin{deluxetable}{lccccc}
\footnotesize
\tablenum{3} 
\tablewidth{0pt} 
\tablecaption{Measured Intensities, Linewidths \& Centroids \label{IntT}} 
\tablehead{ 
\colhead{GMC\tablenotemark{a}} 
&\colhead{Molecule}
&\colhead{T$_{b}$} 
&\colhead{I$_{mol}$} 
&\colhead{$\Delta v$}
&\colhead{$v_{o}$} \\
\colhead{} 
&\colhead{}
&\colhead{$(K)$}
&\colhead{$(K~km~s^{-1})$}
&\colhead{$(km~s^{-1})$}  
&\colhead{$(km~s^{-1})$}
}
\startdata
A &C$_{2}$H&0.27$\pm$0.05 &15$\pm$2& 75$\pm$8.5 & 19$\pm$3.6 \\
  &HNC & 0.50$\pm$0.07  & 22$\pm$3 & 40$\pm$2.2 & 24$\pm$1.0 \\
  &HNCO & $\lsim$0.078  & $\lsim$3.6 & \nodata &\nodata \\
  &HC$_{3}$N & 0.069$\pm$0.02  & 4.1$\pm$0.8 & \nodata &\nodata \\
  &N$_{2}$H$^{+}$ & 0.18$\pm$0.03  & 8.9$\pm$1 & 38$\pm$5.7 &26$\pm$2.4 \\
  &CH$_{3}$OH & 0.16$\pm$0.04  & 6.4$\pm$0.9 & 40$\pm$6.7 &24$\pm$2.9 \\
  &C$^{34}$S & 0.089$\pm$0.02  & 8.5$\pm$1 & 120$\pm$38 &-5.3$\pm$16 \\
B &C$_{2}$H& 0.21$\pm$0.05  & 8.3$\pm$2 & 43$\pm$7.3 &4.8$\pm$3.1 \\
  &HNC & 0.62$\pm$0.09  & 31$\pm$5 & 30$\pm$1.2 &24$\pm$0.6 \\
  &HNCO & $\sim$0.11  & 5.3$\pm$1 & \nodata &\nodata \\
  &HC$_{3}$N & 0.088$\pm$0.02  & 5.4$\pm$0.9 & 20$\pm$9.5 &19$\pm$4.3 \\
  &N$_{2}$H$^{+}$ & 0.19$\pm$0.03  & 7.1$\pm$1 & 23$\pm$3.8 &23$\pm$1.5 \\
  &CH$_{3}$OH & 0.15$\pm$0.04  & 4.5$\pm$0.9 & 25$\pm$5.0 &18$\pm$2.2 \\
  &C$^{34}$S & $\sim$0.053  & 3.6$\pm$0.9 & \nodata &\nodata \\
C &C$_{2}$H& $\sim$0.11  & 7.2$\pm$2 & 30$\pm$9 &33$\pm$3.8 \\
  &HNC & 0.47$\pm$0.07  & 23$\pm$3 & 40$\pm$2.7 &46$\pm$1.1 \\
  &HNCO & 0.23$\pm$0.04  & 8.4$\pm$1 & 30$\pm$6.5 &49$\pm$2.7 \\
  &HC$_{3}$N & 0.20$\pm$0.03  & 8.4$\pm$1 & 30$\pm$5.8 &52$\pm$2.5 \\
  &N$_{2}$H$^{+}$ & 0.18$\pm$0.03  & 7.6$\pm$1 & 35$\pm$6.2 &48$\pm$2.6 \\
  &CH$_{3}$OH & 0.35$\pm$0.05  & 15$\pm$2 & 32$\pm$2.8 &48$\pm$1.2 \\
  &C$^{34}$S & 0.078$\pm$0.02  & 3.6$\pm$0.9 & \nodata &\nodata \\
D &C$_{2}$H& $<$0.11  & $<$1.5 & \nodata &\nodata \\
  &HNC & 0.15$\pm$0.03  & 8.4$\pm$1 & 47$\pm$7.8 &47$\pm$3.3 \\
  &HNCO & 0.16$\pm$0.04  & 3.9$\pm$1 & 28$\pm$5.8 &54$\pm$7.5 \\
  &HC$_{3}$N & $<$0.048  & $<$1.5 & 25$\pm$25 &46$\pm$13 \\
  &N$_{2}$H$^{+}$ & 0.11$\pm$0.03  & 3.2$\pm$0.8 & 35$\pm$8.3 &48$\pm$3.5 \\
  &CH$_{3}$OH & 0.12$\pm$0.04  & 4.1$\pm$0.9 & 38$\pm$6.2 &52$\pm$2.7 \\
  &C$^{34}$S & $\lsim 0.048$  & $\lsim$1.8 & \nodata &\nodata \\
D'&C$_{2}$H& $<$0.11  & $<$1.5 & \nodata &\nodata \\
  &HNC & 0.17$\pm$0.03  & 6.7$\pm$1 & 37$\pm$5 &50$\pm$2 \\
  &HNCO & 0.25$\pm$0.04  & 7.2$\pm$0.8 & 27$\pm$3 &56$\pm$2 \\
  &HC$_{3}$N & $\sim$0.04  & $\sim$1.1 & 26$\pm$14 &53$\pm$6 \\
  &N$_{2}$H$^{+}$ & 0.13$\pm$0.02  & 5.5$\pm$1 & 40$\pm$6 &54$\pm$3 \\
  &CH$_{3}$OH & 0.26$\pm$0.03  & 8.6$\pm$1 & 31$\pm$3 &52$\pm$1 \\
  &C$^{34}$S & $<0.11$  & $<$1.5 & \nodata &\nodata \\
\tablebreak
E &C$_{2}$H& $\lsim$0.11  & 6.0$\pm$2 & 47$\pm$11 &1.0$\pm$4.6 \\
  &HNC & 0.46$\pm$0.07  & 18$\pm$3 & 35$\pm$2.2 &18$\pm$0.9 \\
  &HNCO & 0.25$\pm$0.04  & 5.8$\pm$1 & 22$\pm$3.3 &15$\pm$1.8 \\
  &HC$_{3}$N & 0.084$\pm$0.02  & 3.6$\pm$0.8 & 37$\pm$14 &11$\pm$5.8 \\
  &N$_{2}$H$^{+}$ & 0.21$\pm$0.03  & 6.7$\pm$1 & 28$\pm$4.8 &19$\pm$2.0 \\
  &CH$_{3}$OH & 0.16$\pm$0.04  & 7.3$\pm$1 & 30$\pm$5 &15$\pm$2.1 \\
  &C$^{34}$S & 0.089$\pm$0.02  & 2.7$\pm$0.9 & 25$\pm$5.5 &4.6$\pm$2.5 \\
\enddata
\tablecomments{T$_{b}$ is the main-beam brightness temperature in
units of Kelvins based on the resolutions given in Table \ref{ObsT}.
$I_{mol}$ is the peak integrated intensity in units of K km s$^{-1}$ for
the same resolution.  Uncertainties are based on the larger of the 
RMS noise or an estimated $\simeq 15\%$ absolute calibration errors 
for the temperatures and intensities, and $1\sigma$ from the least-squared 
gaussian fits for the velocity information.  Upper limits represent 
2$\sigma$ values.}
\tablenotetext{a}{Peaks are based on the C$^{18}$O data.  See 
MT01 for coordinates, except for D' which has coordinates 
$\alpha = 03^{h} 46^{m} 49^{s}.8$, $\delta = +68^{o} 05' 59.^{''}2$ 
(J2000).}
\end{deluxetable}

\clearpage

\begin{deluxetable}{lccccccccc} 
\footnotesize
\tablenum{4} 
\tablewidth{0pt} 
\tablecaption{Molecular Abundances in IC 342 \label{AbuT}} 
\tablehead{ 
\colhead{GMC}  
&\colhead{N(H$_{2}$)}
&\colhead{C$_{2}$H}
&\colhead{HNC} 
&\colhead{HNCO}  
&\colhead{HC$_{3}$N}  
&\colhead{N$_{2}$H$^{+}$} 
&\colhead{CH$_{3}$OH}
&\colhead{C$^{34}$S}  
&\colhead{SO\tablenotemark{a}} 
}  
\startdata
A  &2.3 & 3(8) & 2(9) & $<$1(9) & $\lsim$1(9) & 5(10) & 5(9) & 2(9) 
& $<$7(9) \\
B  &2.8 & 1(8) & 2(9) & 2(9) & 1(9) & 3(10) & 3(9) & 6(10) & $<$6(9) \\
C  &3.2 & 1(8) & 1(9) & 2(9) & 2(9) & 3(10) & 8(9) & 5(10) & $<$5(9) \\
D  &1.7 & $<$4(9) & 8(10) & 2(9) & $<$6(10) & 2(10) & 4(9) & $\lsim$5(10) 
& $<$8(9) \\
D' &2.0 & $<$3(9) & 8(10) & 3(9) & $\lsim$3(9) & 5(10) & 8(9) & $<$4(10) 
& $<$8(9) \\
E  &3.5 & $\sim$7(9) & 9(10) & 1(9) & 1(9) & 2(10) & 4(9) & $\sim$3(10) 
& $<$5(9) \\
\enddata
\tablecomments{Format for entries are a(b)= a$\times 10^{-b}$, except
N(H$_{2}$) which is in units of $\times 10^{22}$ cm$^{-2}$.  Each
molecule is based on the assumptions of optically thin line emission
with T$_{x}$ also 10 K.  Upper limits are 2$\sigma$.  Uncertainties
are dominated by systematics and are at least a factor of 3 (see text
for discussion of uncertainties).  H$_{2}$ column densities are based
on C$^{18}$O(1-0) emission sampled at 6$^{''}$ resolution.  T$_{x}$ =
10 K and an abundance of [H$_{2}$/C$^{18}$O] = 2.94$\times 10^{6}$ are
adopted (\S\ref{tempover}).}
\tablenotetext{a}{In determining SO upper limits, a linewidth of 
30 km s$^{-1}$ has been assumed.}
\end{deluxetable}

\clearpage

\begin{deluxetable}{lcccc}
\footnotesize
\tablenum{5} 
\tablewidth{0pt} 
\tablecaption{Other Selected Transitions \label{othermolT}} 
\tablehead{ 
\colhead{Molecule}
&\colhead{Transition}
& \colhead{$\nu$} 
& \colhead{T$_{mb}$} 
& \colhead{GMC} \\  
\colhead{}
&\colhead{}
& \colhead{$(GHz)$} 
& \colhead{$(mJy/bm)$} 
& \colhead{}
}
\startdata
HC$_{5}$N &  33-32 &87.8636 & 22$\pm$7 & B \\
HC$_{5}$N &  35-34 &93.1881 & $<$13 & \nodata \\
HCC$^{13}$CN&  10-9 & 90.6018& $<$10 & \nodata \\
C$_{2}$S & $7_{7}-6_{6}$ &90.6864& $<$10 & \nodata \\
$^{13}$C$^{34}$S & 2-1 &90.9260& 13$\pm$5 & C \\
CH$_{3}$CHO & $5_{-24}-4_{-23}$ E &96.4256& \tablenotemark{a}&
\tablenotemark{a}  \\
HCOOCH$_{3}$ & $8_{45}-8_{36}$ A &96.7092& $<$16 & \nodata \\
NH$_{2}$CHO & $5_{14}-4_{13}$ &109.7535& 22$\pm$8\tablenotemark{b} & E \\
H$_{2}$$^{13}$CO & $3_{12}-2_{11}$ & 219.9085 & $<$90 & \nodata \\
C$^{15}$N & $2-1;\frac{5}{2}-\frac{3}{2};F=3-2$ & 219.734 & $<$90 & 
\nodata \\
\hline
\nodata &\nodata& 93.132(4) & 15$\pm$6 & B,C \\
\nodata&\nodata& 96.431(4) & 19$\pm$7\tablenotemark{a} & B,C \\
\nodata&\nodata& 96.466(4) & 14$\pm$7 & B,C,E \\
\nodata&\nodata& 109.221(4) & 20$\pm$9 & B,E \\
\nodata&\nodata& 109.819(4) & 20$\pm$8 & D \\
\nodata&\nodata& 109.839(4) & 22$\pm$8 & D \\
\enddata
\tablecomments{Upper limits are 2$\sigma$.}
\tablenotetext{a}{Blended with C$^{34}$S(2-1).}
\tablenotetext{b}{Blended with SO$_{2}(17_{513}-18_{414}$).}
\end{deluxetable}

\clearpage

\begin{deluxetable}{lccccccccccc} 
\rotate
\footnotesize
\tablenum{6} 
\tablewidth{0pt} 
\tablecaption{PCA Correlation Matrix\label{PCAcorT}} 
\tablehead{ 
\colhead{Maps}  
&\colhead{$^{12}$CO}
&\colhead{C$^{18}$O}
&\colhead{3MM} 
&\colhead{C$_{2}$H}  
& \colhead{C$^{34}$S}  
&\colhead{CH$_{3}$OH} 
&\colhead{HC$_{3}$N}  
& \colhead{HCN}  
& \colhead{HNC}  
& \colhead{HNCO}  
& \colhead{N$_{2}$H$^{+}$}
}  
\startdata 
$^{12}$CO &1.0&&&&&&&&&& \\
C$^{18}$O &0.82 &1.0&&&&&&&&& \\
3MM &0.65&0.76&1.0&&&&&&&& \\
C$_{2}$H &0.53&0.62&0.76&1.0&&&&&&& \\
C$^{34}$S &0.38&0.39&0.48&0.50&1.0&&&&&& \\
CH$_{3}$OH &0.75&0.80&0.67&0.49&0.21&1.0&&&&& \\
HC$_{3}$N &0.60&0.71&0.85&0.57&0.30&0.68&1.0&&&& \\
HCN\tablenotemark{a} &0.65&0.75&0.90&0.74&0.49&0.64&0.77&1.0&&& \\
HNC &0.76&0.85&0.94&0.79&0.49&0.72&0.81&0.91&1.0& \\
HNCO &0.67&0.75&0.58&0.42&0.19&0.78&0.57&0.57&0.66&1.0& \\
N$_{2}$H$^{+}$ &0.73&0.81&0.75&0.59&0.42&0.71&0.68&0.74&0.82&0.69&1.0 \\
\enddata 
\tablenotetext{a}{Data from \citep[][]{Dow92}.  The data was taken at 
the Plateau de Bure Interferometer and hence has a slightly smaller 
primary beam.}
\end{deluxetable} 
 
\clearpage 
 
\begin{deluxetable}{lccccccc} 
\footnotesize
\tablenum{7} 
\tablewidth{0pt} 
\tablecaption{PCA Eigenvectors \label{PCAT}} 
\tablehead{ 
\colhead{PCA Comp.}  
&\colhead{1}
&\colhead{2}
&\colhead{3} 
&\colhead{4}  
&\colhead{5}  
&\colhead{6} 
&\colhead{7}
}  
\startdata 
$^{12}$CO&0.30 &0.19 &0.36 &0.075 &-0.69 &0.084 &0.27 \\
C$^{18}$O&0.33 &0.16 &0.17 &0.063 &-0.14 &-0.039 &0.009 \\
3MM&0.33 &-0.18 &-0.30 &-0.12 &0.032 &0.032 &0.080 \\ 
C$_{2}$H &0.28 &-0.37 &-0.16 &0.75 &0.064 &0.13 &-0.25 \\
C$^{34}$S &0.18 &-0.62 &0.62 &-0.33 &0.20 &0.21 &0.070 \\
CH$_{3}$OH&0.30 &0.38 &0.076 &0.003 &0.021 &0.44 &-0.56 \\
HC$_{3}$N&0.30 &0.026 &-0.44 &-0.54 &0.034 &0.21 &-0.16 \\
HCN&0.33 &-0.19 &-0.23 &-0.038 &0.024 &-0.082 &0.43 \\
HNC&0.35 &-0.11 &-0.13 &-0.058 &-0.074 &-0.10 &0.18 \\
HNCO &0.27 &0.43 &0.21 &0.11 &0.67 &0.10 &0.39 \\
N$_{2}$H$^{+}$ &0.32 &0.078 &0.15 &-0.089 &0.090 &-0.81 &-0.39 \\
\hline
Egnv. \% &70&10&5.8&3.4&2.7&2.4&1.8 \\
\enddata 
\end{deluxetable} 

\clearpage

\begin{deluxetable}{lcc}
\footnotesize
\tablenum{8} 
\tablewidth{0pt} 
\tablecaption{Selected Intensities Ratios \label{RatT}} 
\tablehead{ 
\colhead{Location} 
& \colhead{$\frac{HCN(1-0)}{HNC(1-0)}$\tablenotemark{a}}
& \colhead{$\frac{SO}{23*C^{34}S}$\tablenotemark{b}}
}
\startdata
A & 2.2$\pm$0.4 & $<$0.15  \\
B & 1.5$\pm$0.3 & $<$0.43   \\
C & 1.6$\pm$0.3 & $<$0.43   \\
D & $\sim$0.99 & $<$0.70    \\
D' & $\sim$1.2 & \nodata    \\
E & 1.6$\pm$0.3 & $<$0.72    \\
\enddata
\tablecomments{The measurements of the HCN/HCN ratio is based on
the resolution of HNC(1-0) given in Table \ref{ObsT}.  The
uncertainties reflect the larger of the absolute calibration
uncertainty or the map noise.  Upper limits are $2\sigma$.}
\tablenotetext{a}{The HCN(1-0) is kindly provided by \citet*[][]{Dow92}.}
\tablenotetext{b}{The SO/CS $abundance$ ratio based on the most
constraining of the SO upper limits.}
\end{deluxetable}

\clearpage 
 
\clearpage 
 
\clearpage 
 
\clearpage 
 
\clearpage 
 
\begin{figure}
\epsscale{0.75}
\plotone{figure5.ps}
\caption{Same as in Figure \ref{spec1} except for the five remaining 
transitions. \label{spec2} }
\end{figure}

\clearpage 
 
\begin{figure}
\epsscale{0.9}
\plotone{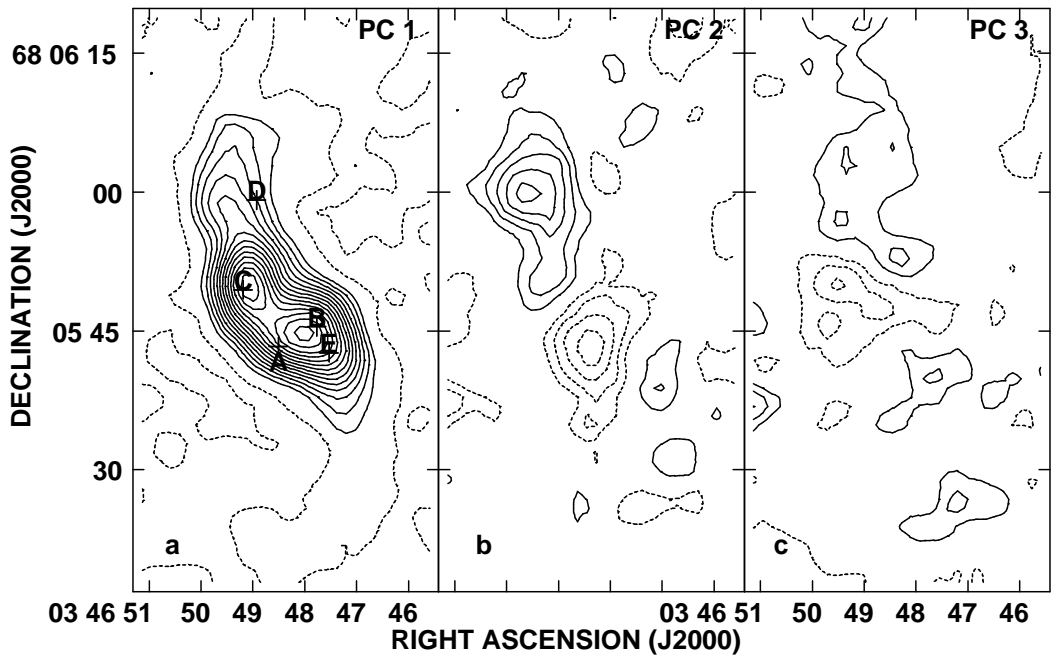}
\caption{The maps of the first three principle components of the 
molecular distribution. \label{pcamap} }
\end{figure}

\clearpage 
 
\begin{figure}
\epsscale{0.9}
\plotone{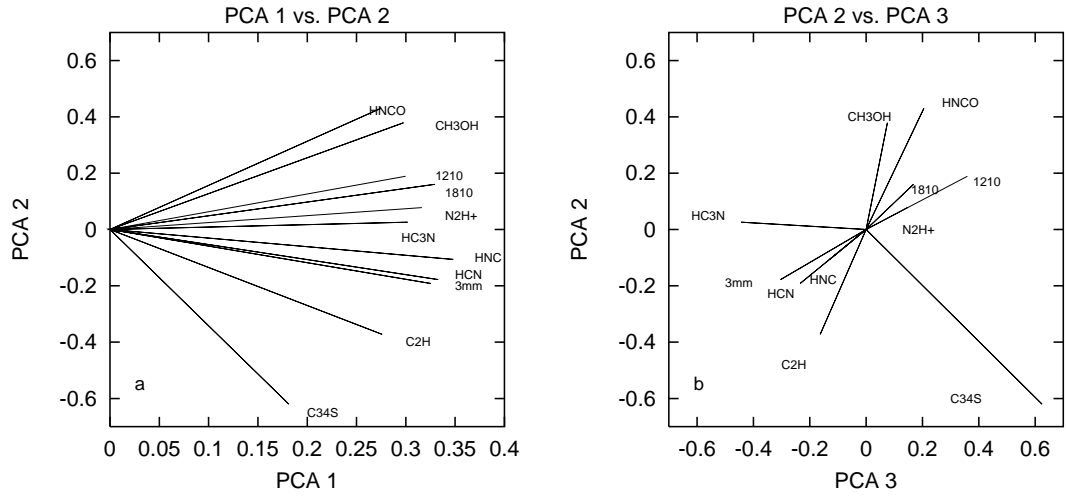}
\caption{The projections of each species on the first three principle
components.  {\it a)} Each transitions projected onto the plane
defined by the first and second principle components.  {\it b)} Each
transitions projected onto the plane defined by the second and third
principle components.  The figure is plotted in such as way that {\it
b)} may be visualized as looking down the $x$-axis of {\it a)} from
the right. \label{pcavect}}
\end{figure}

\clearpage

\begin{figure}
\epsscale{0.9}
\plotone{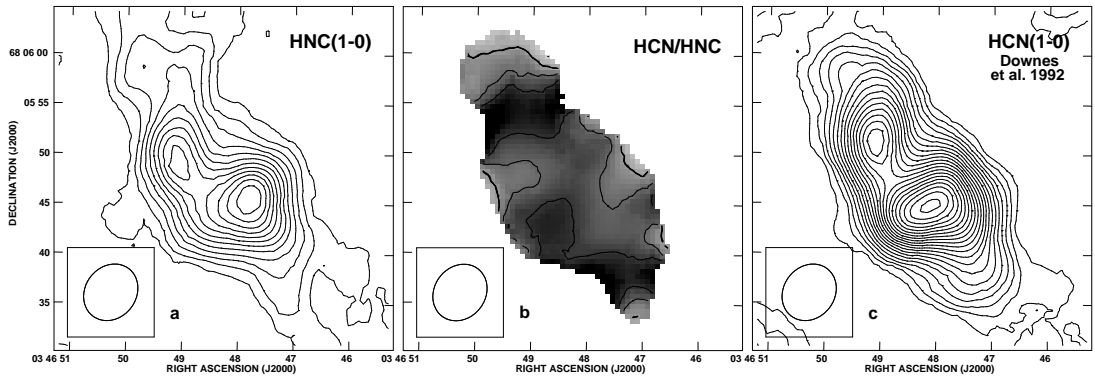}
\caption{ The HCN(1-0)/HNC(1-0) line ratio.  {\it a)} The HNC(1-0)
integrated intensity map in contours of 2.5 K km s$^{-1}$ for the
beamsize given in Table (\ref{ObsT}). {\it b)} The HCN(1-0)/HNC(1-0)
line ratio map towards IC 342.  Contours are 0.5, 1.0, 1.5 and 2.0,
with HCN/HNC = 1.0 contour in bold.  {\it c)} The HCN(1-0) integrated
intensity map of \citet*[][]{Dow92}, convolved to the same resolution,
plotted on the same scale and with the same contours.
\label{hcn/hnc}}
\end{figure}

\clearpage 
 
\begin{figure}
\epsscale{0.9}
\plotone{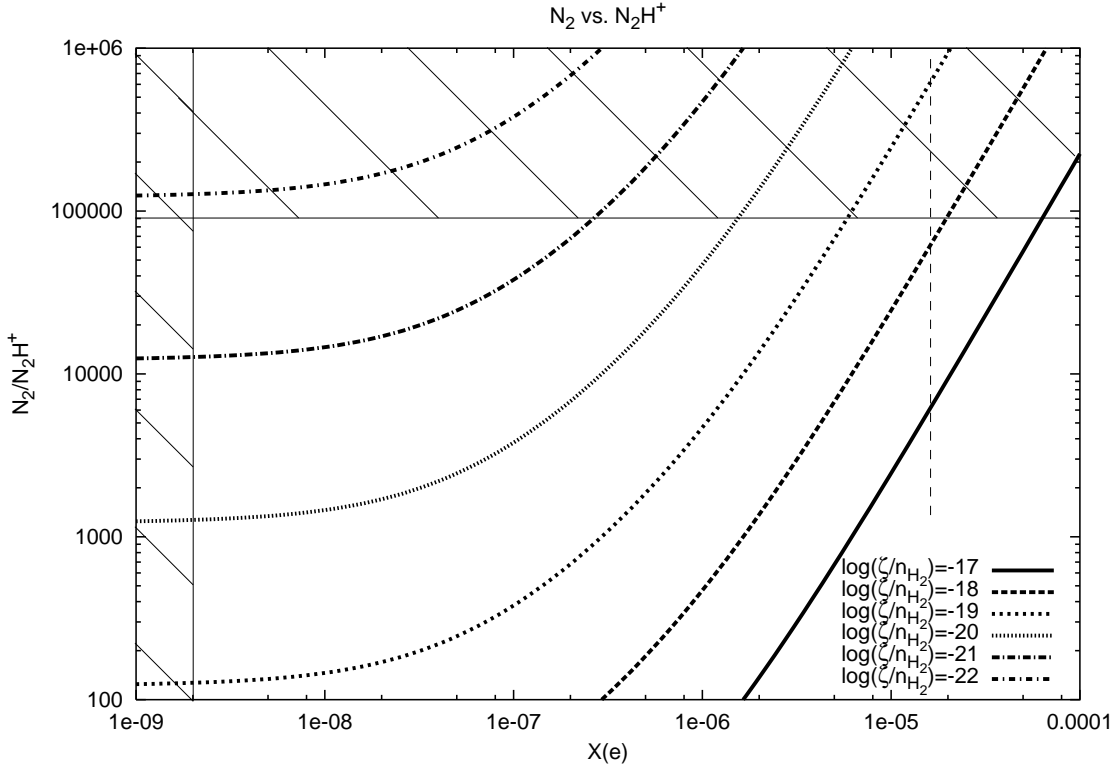}
\caption{The N$_{2}$/N$_{2}$H$^{+}$ abundance ratio as a function of
various physical parameters, based on the basic, steady-state chemical
model described in section \ref{quiesgas}.  The hashed out regions of
$X$$\rm(e^{-})~<~2 \times 10^{-9}$ and N$_{2}$/N$_{2}$H$^{+} ~>~ 9
\times 10^{4}$ are forbidden because the ISM is assumed to be neutral
($X$$(\rm e^{-})$$ ~>~ X$$\rm (HCO^{+}) ~+~ $$X$$\rm (N_{2}H^{+})$) and
that the maximum N$_{2}$ abundance is one half of the cosmic N
abundance, respectively.  The observed $X$$\rm (C^{+})$ ($\simeq ~
X$$\rm (e^{-})$) abundance is also marked {\it (thin dashed line)}.
However this value likely applies to the diffuse molecular gas
component not the dense GMC component traced in N$_{2}$H${+}$.  The
set of thick contours mark the N$_{2}$/N$_{2}$H$^{+}$ - $X$$\rm
(e^{-})$ values for different values of the cosmic ionization rate per
molecular gas density, [$\zeta/n_{H_{2}}$].  Assuming $n_{H_{2}}~>~
10^{4}$ cm$^{-3}$, expected for the excitation of bright
N$_{2}$H$^{+}$, $log(\zeta/n_{H_{2}}) ~\lsim ~ 10^{21}$ corresponds to
a Galactic cosmic ionization rate.  For such an ionization rate,
$X$$\rm (e^{-})~<~10^{-7}$ and N$_{2}~\gsim~ 1 \times 10^{-5}$.  If
$n_{H_{2}}~\gsim ~ 5 \times 10^{-4}$ cm$^{-3}$ in the N$_{2}$H$^{+}$
emitting regions then $\zeta_{IC342}$ must be larger than the Galactic
disk value.  \label{Fn2_n2h}}
\end{figure}

\clearpage 
 
\begin{figure}
\epsscale{0.75}
\plotone{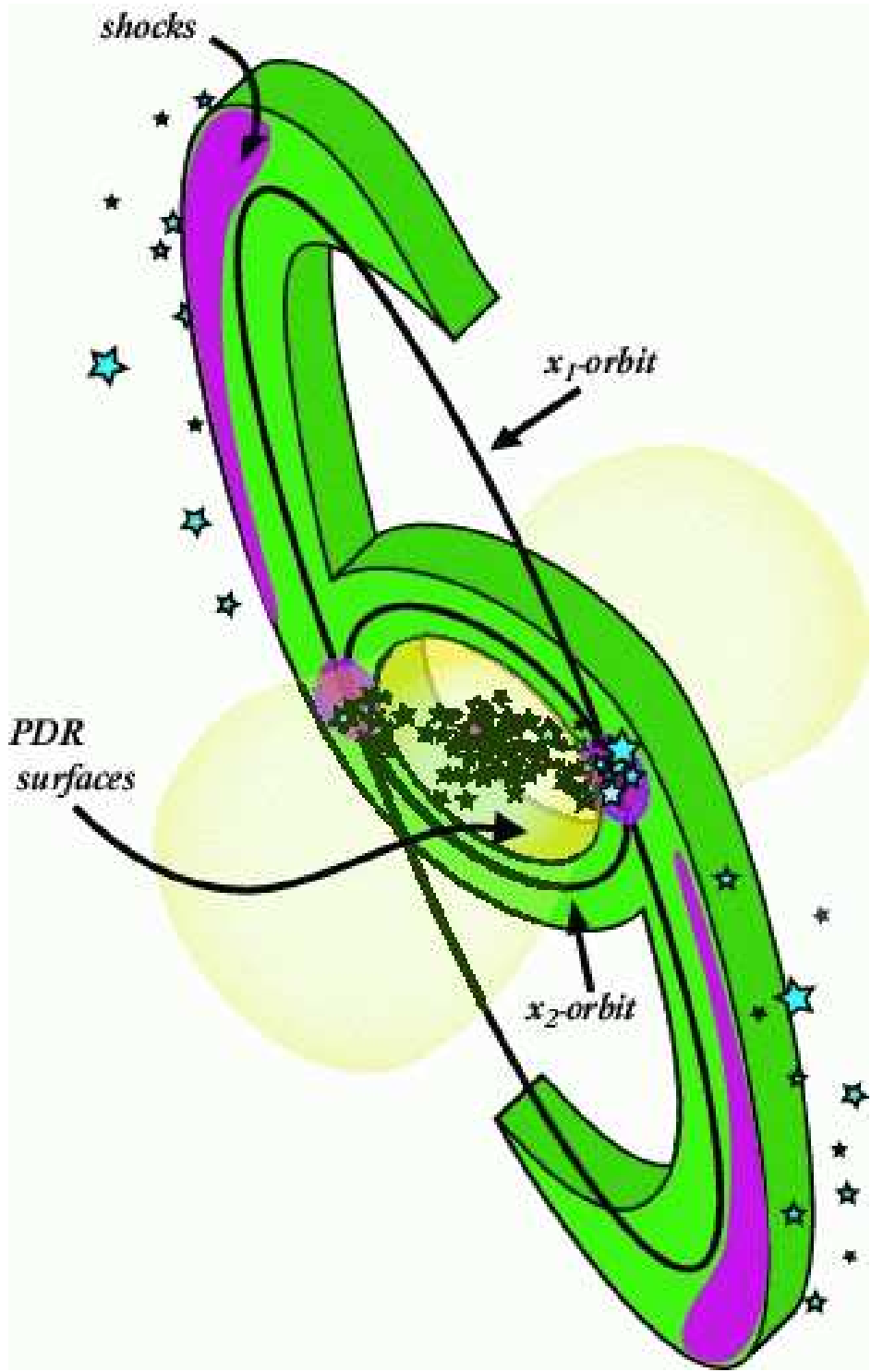}
\caption{Schematic of the chemical and physical structure of the
nucleus of IC 342. \label{schem}}
\end{figure}

\end{document}